\documentclass[aip,jap,amsmath,amssymb,preprint]{revtex4-1}

\usepackage{natbib,amssymb,graphicx,lineno}
\usepackage{rotating}
\usepackage{enumerate}
\usepackage{setspace}
\usepackage{multirow, color}
\usepackage{siunitx}
\usepackage{subcaption}
\usepackage{booktabs,threeparttable}

\begin{document}

\newcommand{\planef}[3]{\ensuremath{\left\{ #1#2#3 \right\} }}
\newcommand{\plane}[3]{\ensuremath{\left( #1#2#3 \right)}}
\newcommand{\dirf}[3]{\ensuremath{\left\langle #1#2#3 \right\rangle}}
\newcommand{\dir}[3]{\ensuremath{\left[ #1#2#3 \right]}}

\title[Fe--He paper]{Binding Energetics of Substitutional and Interstitial Helium and Di-Helium Defects with Grain Boundary Structure in $\alpha$-Fe}

\author{M.A. Tschopp}
\email[Corresponding author, email: ]{mark.tschopp@gatech.edu}
\affiliation{Dynamic Research Corporation, (on site at) U.S. Army Research Laboratory, Aberdeen Proving Ground, MD}
\affiliation{Center for Advanced Vehicular Systems, Mississippi State University, Starkville, MS}
\author{F. Gao}
\author{L. Yang}
\affiliation{Pacific Northwest National Laboratory, Richland, WA}
\author{K.N. Solanki}
\affiliation{Arizona State University, School for Engineering of Matter, Transport and Energy, Tempe, AZ}

\begin{abstract}

The formation/binding energetics and length scales associated with the interaction between He atoms and grain boundaries in BCC $\alpha$-Fe was explored.  Ten different low $\Sigma$ grain boundaries from the \dirf100 and \dirf110 symmetric tilt grain boundary systems were used.  In this work, we then calculated formation/binding energies for 1--2 He atoms in the substitutional and interstitial sites (HeV, He$_2$V, HeInt, He$_2$Int) at all potential grain boundary sites within 15 \AA\ of the boundary (52826 simulations total).  The present results provide detailed information about the interaction energies and length scales of 1--2 He atoms with grain boundaries for the structures examined.  A number of interesting new findings emerge from the present study.  For instance, the $\Sigma3$\plane112 twin boundary in BCC Fe possesses a much smaller binding energy than other boundaries, which corresponds in long time dynamics simulations to the ability of an interstitial He defect to break away from the boundary in simulations on the order of nanoseconds.  Additionally, positive correlations between the calculated formation/binding energies of the He defects ($R >$ 0.9) asserts that the local environment surrounding each site strongly influences the He defect energies and that highly accurate quantum mechanics calculations of lower order defects may be an adequate predictor of higher order defects.  Various metrics to quantify or classify the local environment were compared with the He defect binding energies.   The present work shows that the binding and formation energies for He defects are important for understanding the physics of He diffusion and trapping by grain boundaries, which can be important for modeling He interactions in polycrystalline steels.


\end{abstract}


\maketitle

\section{\label{sec:sec1}Introduction}

Understanding radiation damage phenomena in nuclear materials is important for predicting the mechanical behavior of current and future nuclear power reactors \citep{Ull1984}.  In particular, future fusion reactors will produce a much larger amount of both He and H as compared to fission reactors, hence the microstructure of the structural materials used in fusion reactors will be much more sensitive to interactions with He defects \cite{Blo2007,Zin2009}.  In terms of radiation damage, the production of helium through (n,$\alpha$) transmutation reactions causes both microstructure evolution and drastic property changes in the first-wall and blanket structural materials of fusion reactors.  The production of single helium atoms and small He clusters in the metal lattice is inherently a problem that occurs at the nanoscale.  The subsequent diffusion of He and He clusters results in the nucleation and growth of He bubbles on grain boundaries and within the lattice, which lead to a macroscopic deterioration of material properties including void swelling, surface roughening and blistering, and high temperature intergranular embrittlement \citep[\textit{e.g.},][]{Ull1984,Blo2007,Zin2009,Yam2006,Tri2003,Man1983,Sto1990,Vas1991,Sch2000}.  While the production and diffusion of He occurs at the nanoscale, these other processes develop at larger length scales over long time scales, which necessitates developing predictive multiscale models for material behavior under irradiation conditions that couples multiple simulation methods at different length and time scales.   Developing this predictive capability will require an understanding of the mechanisms associated with radiation damage phenomena, of the He interaction with microstructures, and of the associated uncertainties.  

It is well known that He interactions in Fe play an important role in the mechanical behavior of steel alloys.  There have been a number of quantum mechanics and molecular dynamics simulations that have examined how He and He clusters affect single crystal lattice properties and physical properties in $\alpha$-Fe \citep{Fu2005, Fu2007, Sel2006, Mor2003a,Mor2003c,Gao2011,Hei2006,Hei2007a,Ter2009, Yan2013, Yan2008b, Yan2008c, Yan2007a, Zu2009, Ven2006, Ste2010,Ste2011, Hay2012,Hay2012a, Jus2009}.  For instance, density functional theory (DFT) simulations have been used to show that interstitial He atoms strongly interact with vacancies and can also be trapped by interstitial atoms (binding energy of 0.3 eV) \cite{Fu2005}.  Ventelon, Wirth, and Domain \cite{Ven2006} probed the interactions between He and self-interstitial atoms (SIAs) in $\alpha$-Fe and found strong binding behavior between interstitial He and SIA clusters, which corresponded with the SIA defect strain field.  Other atomistic studies have examined how He and H interact within the single crystal lattice to form complex He--H clusters \cite{Hay2012,Hay2012a} or how He impacts the production of  irradiation-induced point defects in an Fe--Cr matrix \cite{Jus2009}.  Stewart \textit{et al.} \citep{Ste2010,Ste2011} recently used several Fe-He potentials \citep{Jus2008,Sel2007,Wil1972} to show the effect of the interatomic potential on the resulting dynamics of He transport and He clustering in Fe.  Ascertaining the reactions that occur and quantifying their energetics are very important for a fundamental understanding of how point defects, impurities, substitutional atoms, and helium atoms interact in the single crystal lattice of $\alpha$-Fe.  Furthermore, this information is useful for models that explore the kinetics of He diffusion, trapping (clustering), and detrapping (emission), such as rate theory models \cite{Ort2007,Ort2009,Ort2009a, Xu2010}, kinetic Monte Carlo models \cite{Deo2007a,Deo2007b}, and/or phase field models \cite{Hu2009,Zha2012a}.  

Grain boundaries within these alloy systems can also play a significant role in trapping these point defects and atomic species.  Despite this fact, there have been relatively few studies that have focused on He interactions with grain boundaries \cite{Kur2004,Gao2006,Kur2008, Gao2009,Ter2010b,Zha2010i,Zha2012z,Ter2011, Zha2013, Zha2013z}.  These prior works have been significant for understanding the migration paths and mechanisms of He for a few boundaries using the dimer method \cite{Gao2006, Gao2009}, understanding migration of interstitial He in different grain boundaries using molecular dynamics \cite{Ter2010b}, understanding how the grain boundary strength is affected by He \cite{Zha2010i,Zha2012z} or He bubbles \cite{Ter2011}, or understanding the diffusion and stability of He defects in grain boundaries using first principles \cite{Zha2013, Zha2013z}.  For instance, Kurtz and Heinisch \cite{Kur2004} used a Finnis--Sinclair potential (detailed in Morishita et al.~\cite{Mor2003c}) to show that interstitial He was more strongly bound to the grain boundary core than substitutional He.  Kurtz and Heinisch also found that the maximum He binding energy increases linearly with the grain boundary excess free volume, similar to prior work in FCC nickel \cite{Bas1985}.  In subsequent studies, Gao, Heinisch, and Kurtz \cite{Gao2006} found a relationship between the maximum binding energy and grain boundary energy as well.  Additionally, Gao et al.~started to detail the diffusion trajectories of interstitial and substitutional He atoms along a $\Sigma3$ and $\Sigma11$ grain boundary and found that the dimensionality of migration of interstitial He may depend on temperature (e.g., in the $\Sigma3$\plane112 boundary).  Some recent work has utilized first principles to quantify binding strengths of He and He-vacancy clusters at the $\Sigma5$\plane310 symmetric tilt grain boundary \cite{Zha2013z}.  There are still a number of unresolved issues relating to how He interacts with grain boundaries, though.  For instance, these studies often focus on one He atom in interstitial or substitutional sites, but often do not extend to multiple He atoms interacting with grain boundary sites.  Also, atomistic studies often have not examined a wide range of grain boundary structures to understand the influence of macroscopic variables on He interactions.  Moreover, while the highly non-uniform He binding energies in the grain boundary core have been previously pointed out \cite{Kur2004}, relating these to per-atom metrics based on the grain boundary local environment has not been pursued as frequently.  Additionally, while there is an increasing awareness of interatomic potential effects, many of the interatomic potentials used previously for some of these grain boundary studies have been improved upon with updated interatomic potential formulations \cite{Gao2011, Sel2007, Sto2010} and/or more recent quantum mechanics results showing how magnetism affects He defects in $\alpha$-Fe \cite{Sel2005, Sel2006, Zu2009}.  Last, reconciling the molecular statics simulations of He binding to grain boundaries with recent long time dynamics approaches \cite{Hen2001,Ped2009} can shed light on the migration and trapping/detrapping efficiency of these different grain boundaries.

Hence, in the present work, we have focused on how the local grain boundary structure interacts with He atoms and how the local atomic environment at the boundary influences the binding energetics of 1--2 atom substitutional and interstitial He defects.  Recently, Tschopp and colleagues utilized an iterative approach to systematically quantify the interactions between point defects, carbon, and hydrogen with Fe grain boundaries \cite{Tsc2011,Tsc2012a,Rho2013, Sol2013}.  Herein, this approach is extended by using multiple different starting positions, or instantiations, about each site to more precisely probe the formation and binding energy landscape about ten grain boundaries.  In this paper, we present this extended approach as it applies to four He defects and explore how the grain boundary structure affects the interaction of these four He defects using both molecular static and long time dynamic calculations.  Moreover, we have also explored how different per-atom local environment metrics compare with the calculated energies for the different He defects.  A number of interesting new findings emerge from the present study.  For instance, we find that the $\Sigma3$\plane112 twin boundary in BCC Fe possesses a much smaller binding energy than other boundaries, which corresponds in long time dynamics simulations to the ability of an interstitial He defect to break away from the boundary in simulations on the order of nanoseconds at 300 K, in contrast to other boundaries.    Additionally, we find that the calculated formation/binding energies for substitutional He (i.e., He in a monovacancy) correlates well with interstitial He, substitutional He$_2$, and  interstitial He$_2$ energies for the same atomic site, which asserts that the local environment surrounding each site strongly influences the He defect energies and that highly accurate quantum mechanics calculations of lower order defects may be an adequate predictor of higher order defects.   The present work shows that the binding and formation energies for He defects is important for understanding the physics of He diffusion and trapping by grain boundaries, which can be important for modeling helium interactions in polycrystalline steels.


\section{\label{sec:sec2}Methodology}

\subsection{Grain Boundaries}
The interaction between helium and iron grain boundaries was investigated by using ten different grain boundaries and multiple different He defect combinations for multiple sites within 15 \AA\ of the boundary (52826 simulations total).  Table \ref{table1} lists the ten grain boundaries studied, their dimensions in terms of lattice units, the number of atoms and the interfacial energy.  These grain boundaries represent the ten low coincident site lattice (CSL) boundaries ($\Sigma\leq13$) within the \dirf100 and \dirf110 symmetric tilt grain boundary (STGB) systems.  This is a subset of those boundaries used in prior studies of point defect absorption (vacancies and self-interstitial atoms) by a large range of grain boundary structures in pure $\alpha$-Fe \cite{Tsc2011, Tsc2012a}.  

These boundaries were generated using a previous methodology \cite{Tsc2007b,Tsc2007} whereby multiple translations and an atom deletion criteria were used to locate minimum energy grain boundary structures.  This method has been found to agree with experimentally-measured energies for $\Sigma3$ asymmetric tilt grain boundaries in Cu \cite{Tsc2007b, Wol1992} as well as experimental high resolution transmission electron microscopy (HRTEM) images \cite{Ern1996}, including the orthorhombic 9R phase in FCC metals \cite{Ern1992,Hof1994,Ern1996}.  While this methodology for generating grain boundaries was first applied to FCC metals, this method will also work for other BCC and HCP grain boundaries as it is based on generating a large number of energy-minimized grain boundary structures to find the minimum energy grain boundary for each grain boundary.

The current set of boundaries includes four \dirf100 STGBs ($\Sigma5$,$\Sigma13$) and six \dirf110 STGBs ($\Sigma3$,$\Sigma9$,$\Sigma11$).  Recent experimental characterization of steels has shown that several of these symmetric tilt grain boundaries are observed at a concentration higher than random grain boundaries \cite{Bel2013a, Bel2013b}.  For example, Beladi and Rollett quantified that the $\Sigma3$\plane112 symmetric tilt grain boundary is observed at $>$10 multiples of a random distribution (MRD) of grain boundaries \cite{Bel2013a, Bel2013b}, i.e., much larger than would be expected.  While the experimental observation of \dirf100 symmetric tilt grain boundaries ($\Sigma5$, $\Sigma13$ GBs) is below 1 MRD, these grain boundaries are commonly used in DFT studies due to the low periodic distances required in the grain boundary plane.  The present set of boundaries is smaller than those previously explored \cite{Tsc2011, Tsc2012a} for two reasons.  First, since we explored multiple starting configurations for the He defects in this study, a larger number of simulations were required for each grain boundary than for the point defect studies, which only considered a single vacancy or self-interstitial atom.  Second, our prior study \cite{Tsc2012a} found that, aside from a few boundaries (e.g., the $\Sigma3$\plane112 STGB, included herein), most grain boundaries had similar characteristics with respect to point defect interactions.  These results suggest that the ten boundaries explored within can supply ample information about the interaction of He defects with low-$\Sigma$ grain boundaries, and perhaps shed insight on general high angle grain boundaries as well.  

The simulation cell consisted of a 3D periodic bicrystalline structure with two periodic grain boundaries, similar to prior grain boundary studies \cite{Rit1996,Spe2007,Tsc2007}.  The two mirror-image grain boundaries are separated by a minimum distance of 12 nm to eliminate any effects on energies due to the presence of the second boundary.  While the grain boundaries were generated using the minimum periodic length in the grain boundary period direction and the grain boundary tilt direction ($x$- and $z$- directions, respectively), it was found that the formation energies for the defects were influenced for periodic lengths below 4$a_0$.  That is, the periodic image of the defect and/or its influence on the surrounding lattice can significantly affect the defect's formation energy.  Hence, multiple replications in the grain boundary tilt direction and the grain boundary period direction were used.  For instance, the final dimensions for the $\Sigma5$\plane210 GB resulted in a vacancy formation energy far away from the boundary that was within 0.015\% of that within a 2000-atom BCC single crystal (i.e., 10$a_0$ per side).  This criteria resulted in simulation cell sizes on the order of 4660--9152 atoms ($\Sigma{13}$\plane510 and $\Sigma{11}$\plane113, respectively).  All of the simulations were performed with a modified version of the MOLDY code \cite{Ref2000,Ack2011,Gao1997}.

\begin{table}
\centering
\caption{\label{table1} Dimensions of the bicrystalline simulation cells used in this work along with the $\Sigma$ value of the boundary, the grain boundary plane (normal to the $y$-direction), the misorientation angle $\theta$ about the corresponding tilt axis ($z$-direction), the grain boundary energy, and the number of atoms.  The cell dimensions were chosen to ensure convergence of the formation and binding energies of the inserted He defects.  }
\begin{scriptsize}
\begin{ruledtabular}
\begin{tabular}{ccccccccc}
Sigma \&  & GB tilt & $\theta$ & GB energy & $x$ & $y$ & $z$ & Number of & Free Volume \\
GB plane &  direction & ($^\circ$) & (\si{mJ.m^{-2}}) & (\si{\angstrom}) & (\si{nm}) & (\si{\angstrom}) & atoms &  (\AA$^3$/\AA$^2$)\\
\hline \\ [-1.5ex]
$\Sigma3$\plane111 & \dirf110 & 109.47$^\circ$ & \SI{1308} & \SI{21.0}  & \SI{24.8}  & \SI{16.2}  & \num{7200} & 0.35 \\
$\Sigma3$\plane112 & \dirf110 & 70.53$^\circ$ & \SI{260} &  \SI{14.8} & \SI{25.2}  &  \SI{16.2} & \num{5184} & 0.01 \\
$\Sigma5$\plane210 & \dirf100 & 53.13$^\circ$ &   \SI{1113} &  \SI{19.2} &  \SI{24.5} & \SI{14.3} & \num{5730} & 0.35 \\
$\Sigma5$\plane310 & \dirf100 & 36.87$^\circ$ &    \SI{1008} & \SI{18.1}  &  \SI{25.3} & \SI{14.3} &\num{5600} & 0.30 \\
$\Sigma9$\plane221 & \dirf110 & 141.06$^\circ$ &    \SI{1172} & \SI{17.1}  &  \SI{24.2} & \SI{16.2} &\num{5728} & 0.19 \\
$\Sigma9$\plane114 & \dirf110 & 38.94$^\circ$ &   \SI{1286} & \SI{24.2}  & \SI{25.5}  & \SI{16.2} &\num{8576} & 0.35 \\
$\Sigma11$\plane113 & \dirf110 & 50.48$^\circ$ & \SI{1113} & \SI{26.8}  & \SI{24.7}  & \SI{16.2}  &  \num{9152} & 0.26 \\
$\Sigma11$\plane332 & \dirf110 & 129.52$^\circ$ & \SI{1020} &  \SI{18.9} & \SI{24.1}  &  \SI{16.2} &  \num{6336} & 0.21 \\
$\Sigma13$\plane510 & \dirf100 & 22.62$^\circ$ & \SI{1005} &  \SI{14.6} & \SI{26.1}  & \SI{14.3}  &  \num{4660} & 0.27 \\
$\Sigma13$\plane320 & \dirf100 & 67.38$^\circ$ & \SI{1108} &  \SI{20.6} & \SI{24.7}  & \SI{14.3}  &  \num{6220} & 0.23 \\
\end{tabular}
\end{ruledtabular}
\end{scriptsize}
\end{table}

Table \ref{table1} also lists several properties of the ten grain boundaries.  First, notice that the grain boundary energies range from 260--1308 \si{mJ.m^{-2}}, although the majority of the CSL boundaries have energies $>$1000 \si{mJ.m^{-2}}.  Also, all boundaries are high angle grain boundaries, based on a 15$^\circ$ Brandon criterion for low/high angle grain boundaries.  Additionally, note that while the misorientation angles $\theta$ refer to the conventional misorientation angle-energy relationships (e.g., in Ref.~\onlinecite{Tsc2012a}), the disorientation angle, or minimum angle to rotate lattice A to lattice B, is the same for the two instances of each $\Sigma$ boundary.  The misorientation angles are based on deviation from the\plane100 planes in the \dirf100 and \dirf110 STGB systems.  The grain boundary energies are similar to those previously calculated (e.g., $\Sigma5$\plane310 and $\Sigma13$\plane320 GBs are within 2\% and 7\%, respectively, of a prior study \cite{Li2013}).  The grain boundary structures vary for the ten grain boundaries.  Further details on the grain boundary structure are given in Tschopp et al.~\cite{Tsc2012a}.  The grain boundary structures have been compared with computed structures using quantum mechanics, when possible.  For instance, Bhattacharya et al.~used DFT to calculate grain boundary structures for $\Sigma3$\plane111 and $\Sigma11$\plane332 GBs \cite{Bha2013}, which are identical to those computed in the present work.  Moreover, the relationship between the grain boundary energies and excess free volume for the $\Sigma3$\plane111 and $\Sigma11$\plane332 GBs also agrees with previous studies \cite{Shi2009,Bha2013,Kur2004}, as well as with other studies that have found that the $\Sigma3$\plane112 GB has a much lower grain boundary energy and excess free volume in comparison to the $\Sigma3$\plane111 GB \cite{Shi2009,Tsc2012a}.  Additionally, the $\Sigma5$\plane310 and $\Sigma9$\plane114 GB structures also agree with previously calculated first principles structures \cite{Zha2013, Zha2013z}.  Also included in this table is the excess free volume, which was calculated using a previous methodology for calculating excess volume \cite{Wol1991, Tsc2007b} whereby the volume occupied by the bicrystal simulation cell is compared to an equivalent volume of a perfect single crystal lattice and divided by the total grain boundary area.  

\subsection{Interatomic Potential}

The Fe--He interatomic potential fitted by Gao et al.~\cite{Gao2011} to ab initio calculations using an s-band model was used in the present atomistic modeling.  This interatomic potential is based on the electronic hybridization between Fe $d$-electrons and He $s$-electrons to describe the Fe--He interaction.  The single element potentials utilized in the formulation of this potential are the Ackland and Mendelev (AM) potential for the Fe--Fe interactions \cite{Ack2004} and the Aziz et al.~Hartree--Fock-dispersion pair potential (Aziz-potential) \cite{Azi1995} for the He--He interactions.  The atomic configurations and formation energies of both single He defects (substitutional, tetrahedral, and octahedral He) and small interstitial He clusters (He$_2$V, He$_3$V, and He--He di-interstitial) were utilized in the fitting process.  Calculations using this interatomic potential show that both tetrahedral and octahedral interstitials are stable, with tetrahedral He being the most stable interstitial configuration \cite{Gao2011}, which agrees with previous ab initio calculations \cite{Sel2005, Zu2009}.  The binding properties of the He$_x$V and He$_x$ interstitial clusters are in reasonable agreement with ab initio and previous potential results.  This potential has been previously used to investigate the emission of self-interstitial atoms from small He clusters in the $\alpha$-Fe matrix and to show the dissociation of a di-interstitial He cluster at temperatures $>$ 400 K.  The aforementioned potential is deemed appropriate for studying the He interaction with grain boundaries in this work.  In addition, the recent first principles calculations of energetic landscape and diffusion of He in $\alpha$-Fe grain boundaries demonstrate that the potentials used in the present study satisfactorily describe the He behavior at the GBs \cite{Zha2013}.  In fact, this study has shown that there is good agreement between the vacancy formation energies for $n$ vacancies ($n\le{4}$) in the $\Sigma3$\plane111 GB between DFT results and the present empirical potential \cite{Zha2013}.  Furthermore, the present interatomic potential has shown good agreement with DFT results for interstitial and substitutional formation energies for multiple layers and multiple boundaries \cite{Zha2013}.  

\begin{table}
\centering
\caption{\label{table2} Formation energies for the Fe--He potential \cite{Gao2011} used in the present work compared to DFT results  \cite{Sel2005,Sel2006,Sel2007,Fu2007}.}
\begin{scriptsize}
\begin{ruledtabular}
\begin{tabular}{lccccc}
&
	\includegraphics[height=0.15\columnwidth,angle=0]{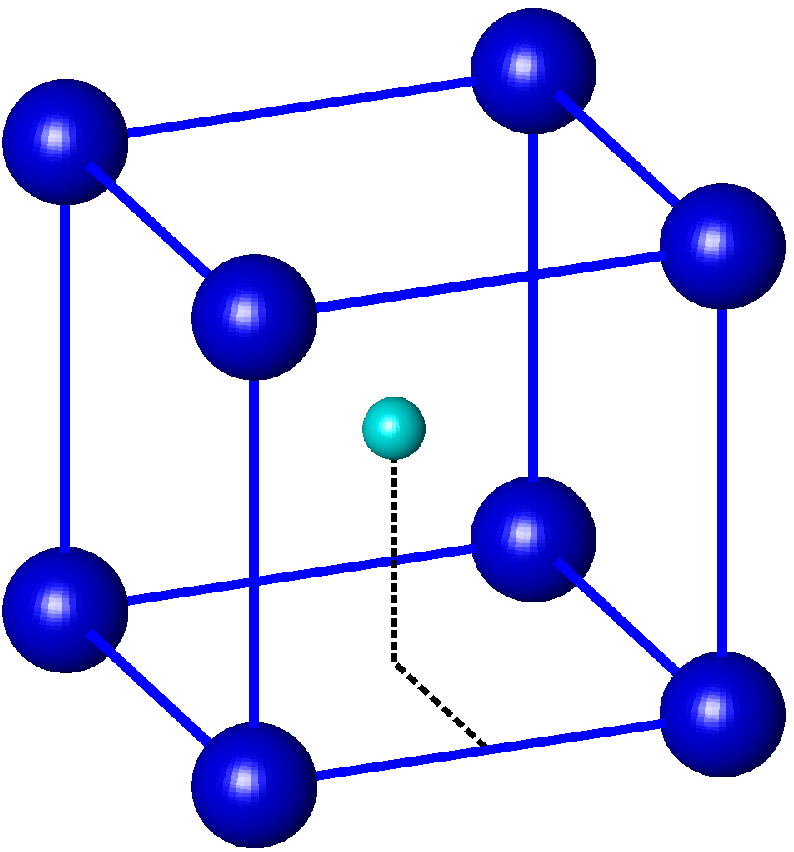} & 
	\includegraphics[height=0.15\columnwidth,angle=0]{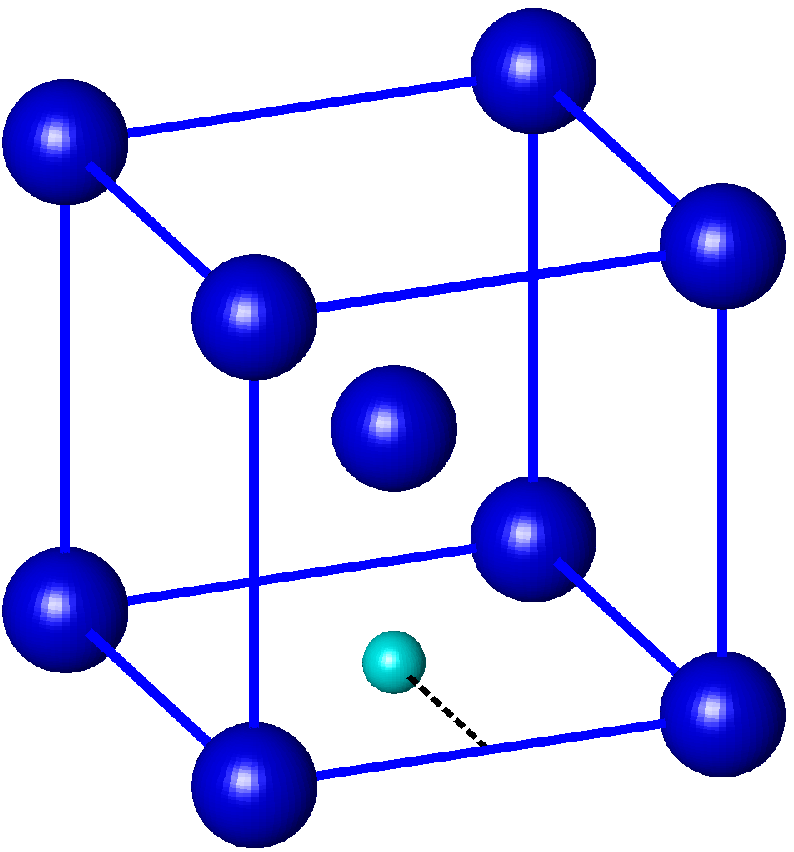} & 
	\includegraphics[height=0.15\columnwidth,angle=0]{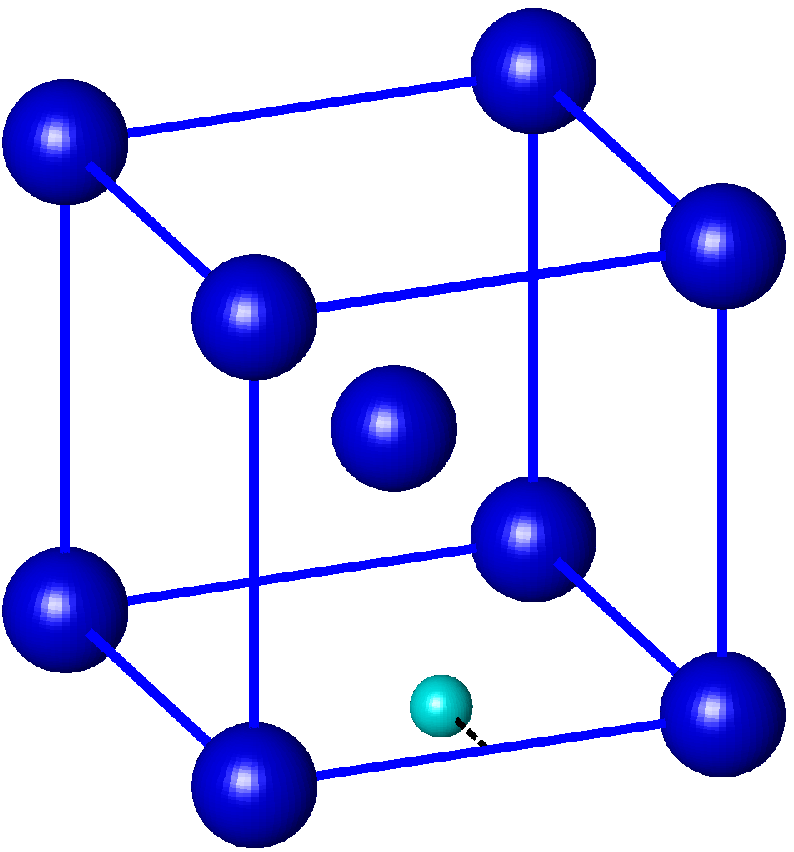} & 
	\includegraphics[height=0.15\columnwidth,angle=0]{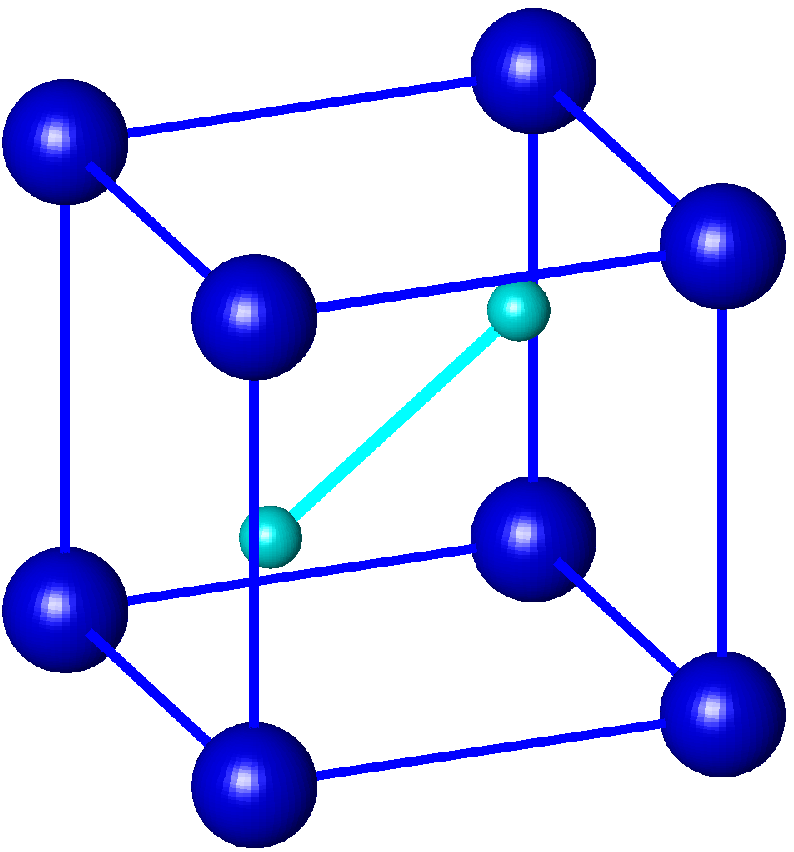} & 
	\includegraphics[height=0.15\columnwidth,angle=0]{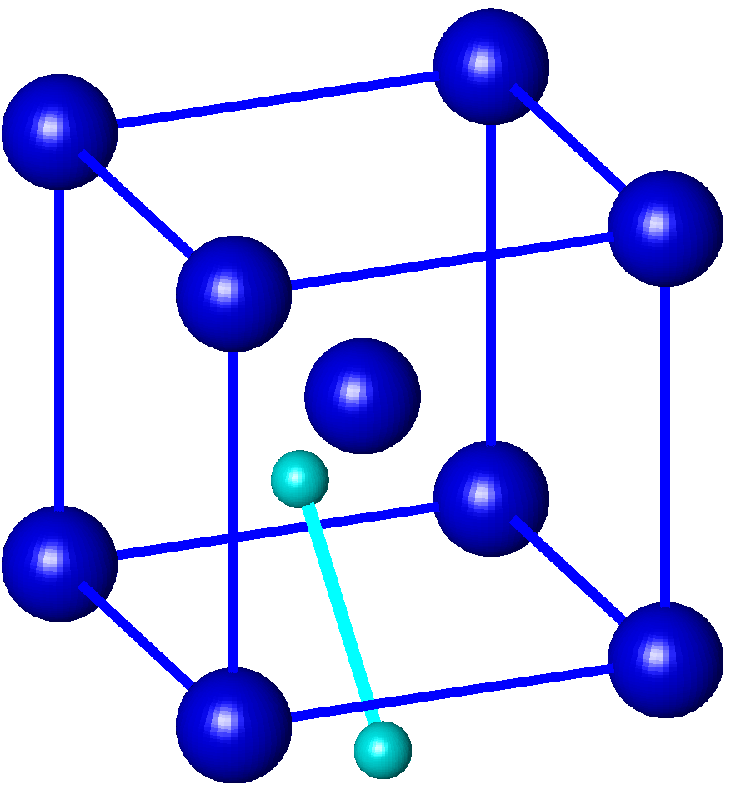} \\ [1ex]
Defect & $HeV$ & $He_{oct}$  & $He_{tet}$ & $He_{2}V$ & $He_{2}$ \\
\hline \\ [-1.5ex]
VASP \cite{Sel2007} & $4.08\left(3.73\right)$ eV\tnote{a} & $4.60$ eV  & $4.37$ eV & $6.63$ eV & $8.72$ eV \\
SIESTA \cite{Fu2007} & 4.22 eV & 4.58 eV & 4.39 eV & -- & -- \\
MD \cite{Gao2011}  & $3.76$ eV\tnote{b} & $4.47$ eV\tnote{b}  & $4.38$ eV\tnote{b} & $6.87$ eV\tnote{b} & $8.49$ eV\tnote{b} \\  
\end{tabular}
        \begin{tablenotes}
		\scriptsize
            	\item [a] The data in parentheses were adjusted by Seletskaia et al.~\cite{Sel2007} for their empirical potential fitting.
            	\item [b] The calculated He formation energies are in agreement with previous results \cite{Gao2011}.
        \end{tablenotes}
\end{ruledtabular}
\end{scriptsize}
\end{table}

\subsection{Helium Defects}

There are four different defect types associated with He atoms that were explored in the present study.  These He-defects included substitutional He (HeV), He interstitial (He$_{int}$), substitutional He$_2$ (He$_2$V), and di-He interstitial (He$_{2,int}$), where the V represents a vacancy.  The substitutional He defect types result from removing an Fe atom and placing either a single He atom or a He--He dumbbell in the empty site.  For substitutional He (HeV), the He atom was simply added in the exact location of the removed Fe atom.  

\begin{figure}[ht!]
  \centering
        \begin{subfigure}[b]{0.475\textwidth}
  \centering
	 \includegraphics[height=2.8in,angle=0]{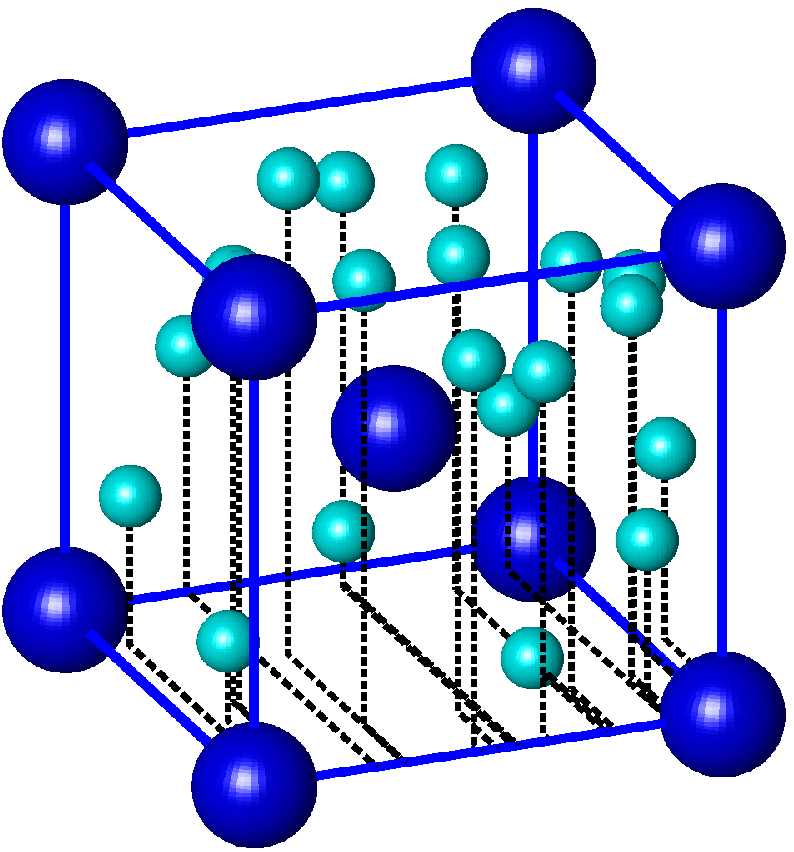}
                \caption{He Interstitial}
                \label{fig0a}
        \end{subfigure}%
        \begin{subfigure}[b]{0.475\textwidth}
  \centering
	 \includegraphics[height=2.8in,angle=0]{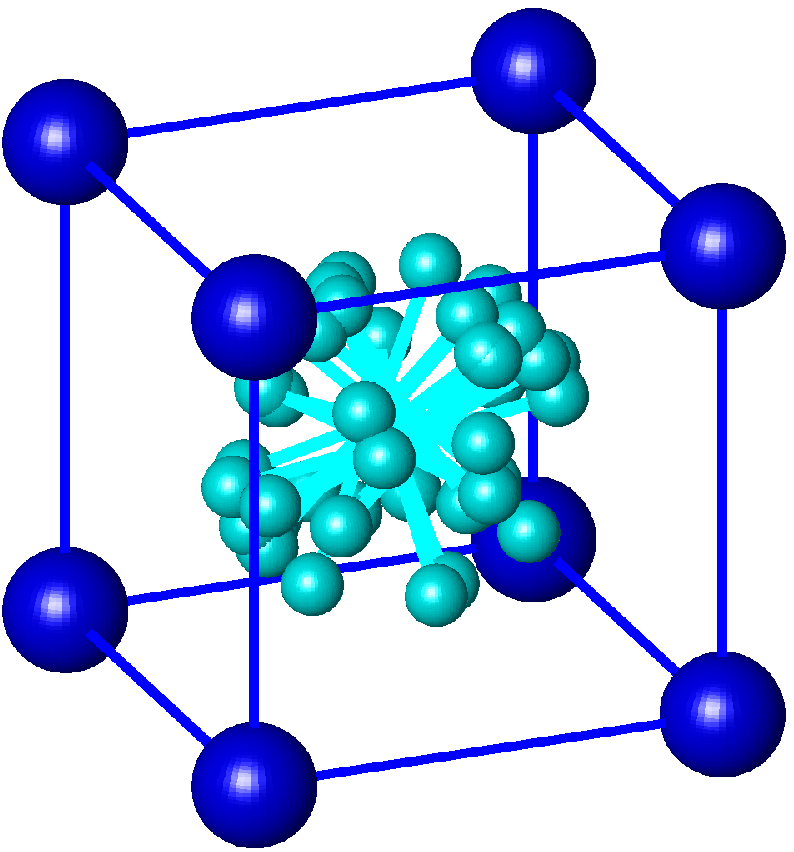}
                \caption{Substitutional He$_2$}
                \label{fig0b}
        \end{subfigure}
\caption{ \label{figure1} An example of the twenty instantiations used for (a) interstitial He atoms and (b) substitutional He$_2$ dumbbells.  }
\end{figure}

For multiple atoms or atoms that are in off-lattice positions, a slightly different methodology was used.  In the case of He$_2$V, the two atoms were placed in opposite directions along a randomly-oriented vector emanating from the vacant site with equal distances to the vacant site and a total distance of 1 \AA.  Since a single instance may not obtain the minimum energy dumbbell, twenty different instances of the starting configurations were used for each potential site for the He$_2$V defects.  In a similar manner, 20 different starting positions were used for He interstitial atoms, which were placed along a randomly-oriented vector emanating from each Fe atom at a distance of $\sqrt{5}/4a_0$.  The He$_2$ interstitial atoms utilized this process for placing the first He atom and then placed the second atom 1 \AA\ away from the first He atom, with a criterion restricting it from placing the atom closer than 1 \AA\ from a neighboring Fe atom.  This number of instances (20) was sufficient to obtain a near constant mean formation energy for interstitial He atom (maximum deviation of 0.4\% of bulk value, mean deviation of 0.03\% of bulk value) in the bulk region far away from the grain boundary.  The main reason for not computing the local crystallography and choosing a predetermined dumbbell orientation for each site in agreement with bulk, for instance, is that other dumbbell orientations may be favored in the grain boundary region and we did not want any assumptions of dumbbell orientation to potentially bias the results.  Figure \ref{figure1} is an example of the 20 instances of interstitial He and He$_2$V surrounding the central atom site.   

\subsection{Formation and Binding Energies}

The formation energies for the He defects could then be calculated as a function of spatial location of sites and their proximity to the grain boundary.    The formation energy for a He defect containing $p$ He atoms and $q$ vacancies at site $\alpha$ of a grain boundary configuration is given by

\begin{equation}
  \label{eq:eq1}
	E^{He_pV_q,\alpha}_{f}=\left(E^{He_pV_q,\alpha}_{tot} + qE_c^{Fe}\right) - \left(E^{GB}_{tot}\right).	
\end{equation}

\noindent Here, $E^{He_pV_q,\alpha}_{tot}$ is the total energy of the grain boundary configuration with the $He_pV_q$ defect at site $\alpha$, $E^{GB}_{tot}$ is the total energy of the grain boundary without any defects, and $E_c^{Fe}$ is the cohesive energy of BCC Fe ($E_c^{Fe}=4.013$ eV).  The cohesive energy of He is negligible and not included in Equation \ref{eq:eq1}.  
 
The binding energy of the He defects with the grain boundary is also of interest.  The total binding energy of a He defect interacting with the GB can be directly calculated from the formation energies of the He defect in the bulk and the He defect at the GB.  As an example of a HeV defect binding to the grain boundary, the binding energy for a HeV defect at site $\alpha$ is given by

\begin{equation}
  \label{eq:eq2}
	E^{HeV^\alpha}_{b}=E^{HeV, bulk}_{f}-E^{HeV^\alpha}_{f},	
\end{equation}

\noindent where $E^{HeV, bulk}_{f}$ and $E^{HeV^\alpha}_{f}$ are the formation energies of a HeV defect either in the bulk or at site $\alpha$, respectively.  It can be seen that a positive binding energy represents that it is energetically favorable for the He defect to segregate to the GB, while a negative binding energy represents that the He defect does not want to segregate to the GB.

\section{Results and Discussion}

\subsection{Spatial distribution of binding energies}

\begin{figure}[b!]
  \centering
	\begin{tabular}{ccccc}
	 &  &  & & \footnotesize $E^b_{HeV}$ (eV) \\
	 \includegraphics[trim = 1.3in 0.0in 1.3in 0.0in, clip, height=3in,angle=0]{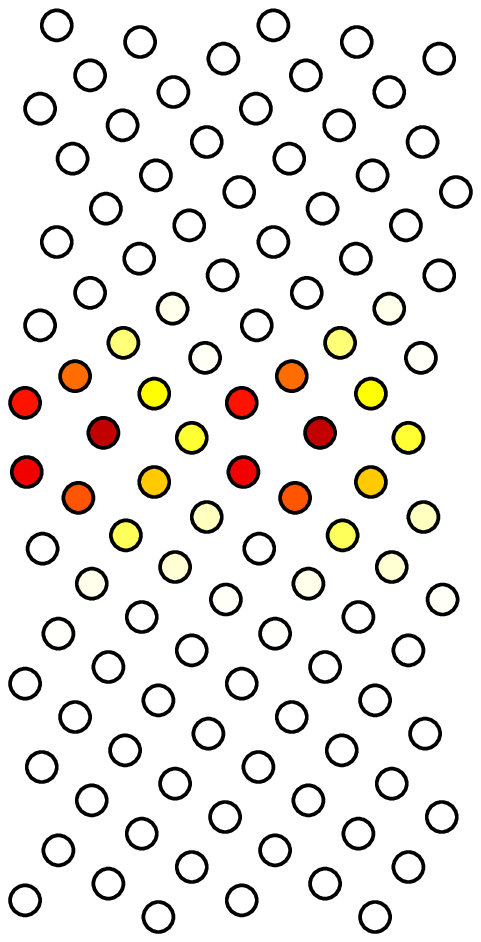} &
	 \includegraphics[trim = 1.7in 0.0in 1.7in 0.0in, clip, height=3in,angle=0]{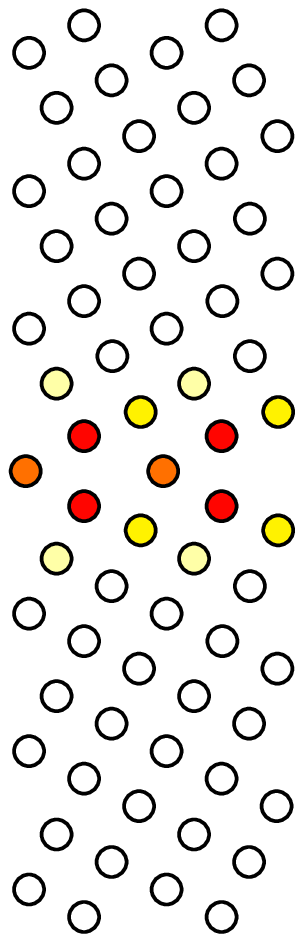} &
	 \includegraphics[trim = 1.8in 0.0in 1.8in 0.0in, clip, height=3in,angle=0]{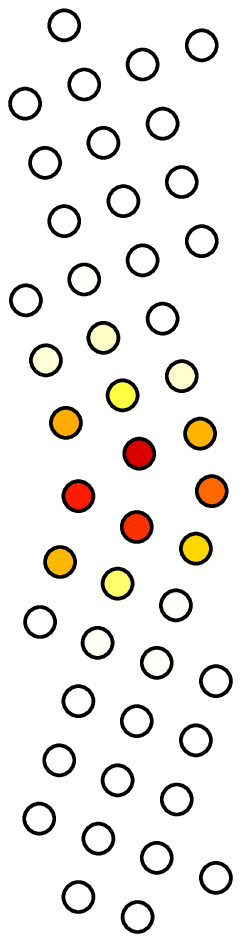} &
	 \includegraphics[trim = 1.6in 0.0in 1.6in 0.0in, clip, height=3in,angle=0]{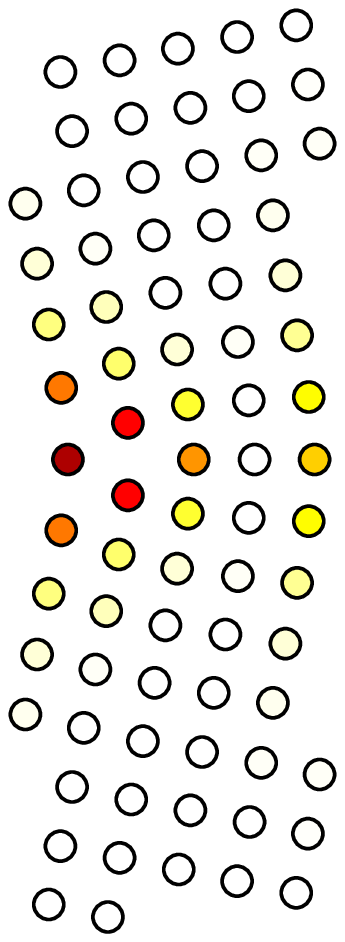} &
	 \includegraphics[trim = 4.0in 0.0in 0.0in 0.1in, clip, height=3in,angle=0]{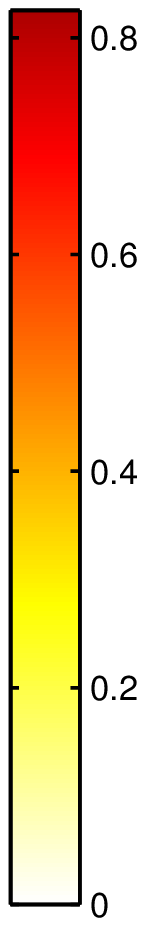} \\
	$\Sigma13$\plane510 & $\Sigma5$\plane310 & $\Sigma5$\plane210 & $\Sigma13$\plane320 & \\
	22.62$^\circ$ & 36.87$^\circ$ & 53.13$^\circ$ & 67.38$^\circ$ &
	\end{tabular}
\caption{ \label{figure2} The binding energies of HeV defect in various sites for the four \dirf100 symmetric tilt grain boundaries.  }
\end{figure}

Figure \ref{figure2} displays the spatial distribution of binding energies for HeV defects in the \dirf100 symmetric tilt grain boundary system.  The grain boundaries in this figure (and subsequent figures) are arranged in order of increasing misorientation angle.  In Figure \ref{figure2}, each atom represents a site where HeV was placed and the binding energy was calculated (i.e., each atom is a different simulation).    The minimum periodic length for each grain boundary is shown along the horizontal axis and the length from top to bottom is 30 \AA, with the grain boundary plane centered in the vertical direction.  The binding energy scale shown to the right of Figure \ref{figure2} ranges from 0 eV (bulk lattice) to the maximum calculated binding energy (0.8256 eV for the $\Sigma13$\plane320 GB) from all ten boundaries.  The atoms far away from the boundary are white (binding energy of 0 eV), indicating that there is no energy difference over the bulk lattice.  As the He defects get closer to the grain boundary, there is a GB-affected region with an increased binding energy for the He defects.  The largest binding energies tend to be along the center of the grain boundary plane or along the 1$^{st}$ layer from the GB plane.  Interestingly, there is a range of binding energies along the central GB plane, as evidenced in the $\Sigma13$\plane320 GB, which contains HeV defect positions with both the largest binding energy and a binding energy similar to bulk within the central grain boundary plane.  Furthermore, there is a noticeable symmetry to the binding energies about the grain boundary plane due to the symmetric nature of these grain boundary structure.

\begin{figure}[b!]
  \centering
	\begin{tabular}{ccccccc}
	 &  &  & & & & \footnotesize $E^b_{HeV}$ (eV) \\
	 \includegraphics[trim = 1.7in 0.0in 1.7in 0.0in, clip, height=2in,angle=0]{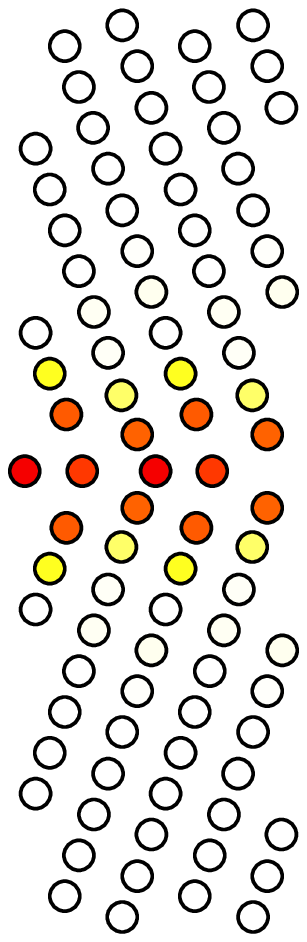} &
	 \includegraphics[trim = 1.4in 0.0in 1.5in 0.0in, clip, height=2in,angle=0]{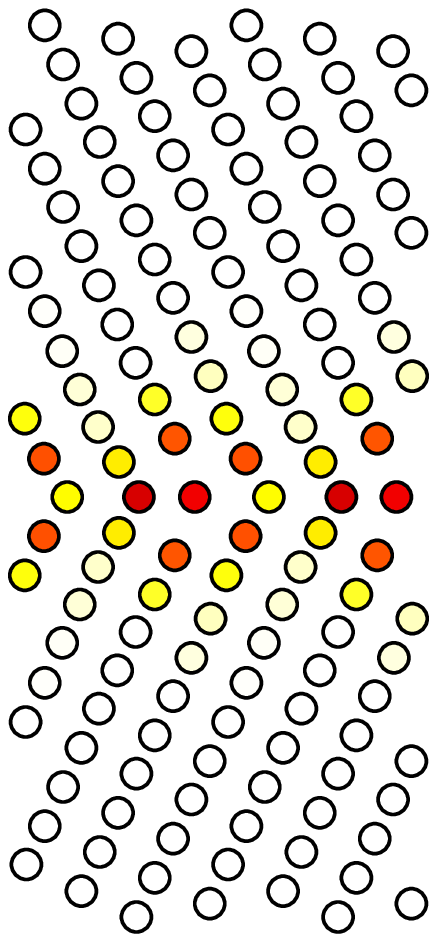} &
	 \includegraphics[trim = 1.9in 0.0in 2in 0.0in, clip, height=2in,angle=0]{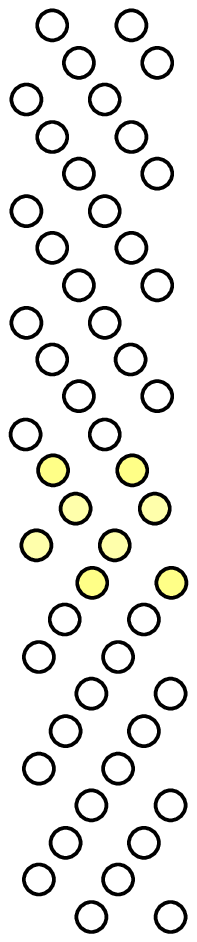} &
	 \includegraphics[trim = 1.8in 0.0in 1.8in 0.0in, clip, height=2in,angle=0]{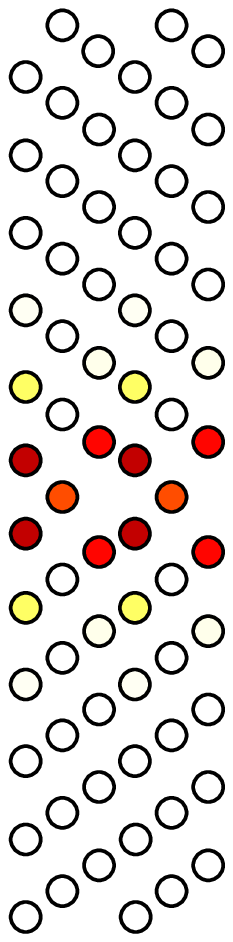} &
	 \includegraphics[trim = 1.6in 0.0in 1.6in 0.0in, clip, height=2in,angle=0]{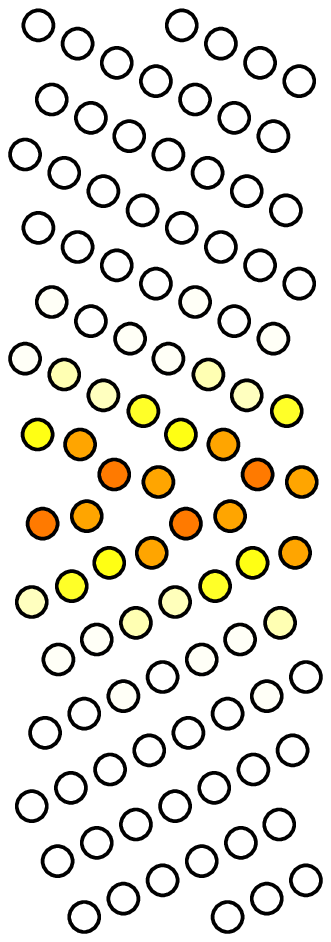} &
	 \includegraphics[trim = 1.5in 0.0in 1.5in 0.0in, clip, height=2in,angle=0]{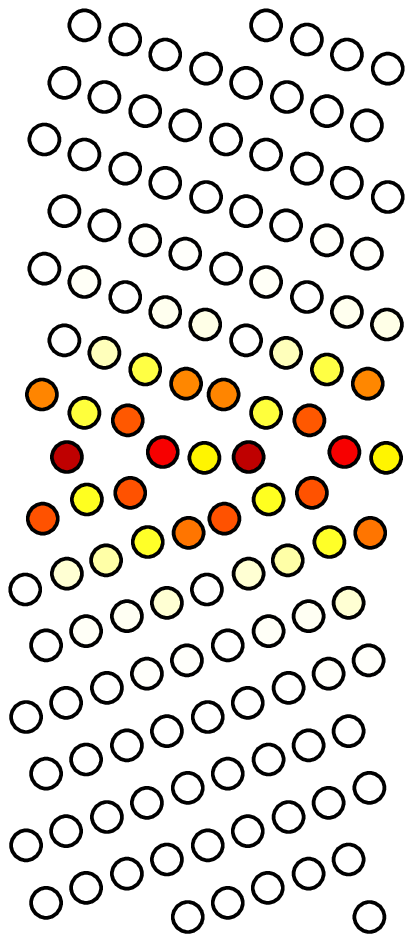} &
	 \includegraphics[trim = 4.0in 0.0in 0.0in 0.1in, clip, height=2in,angle=0]{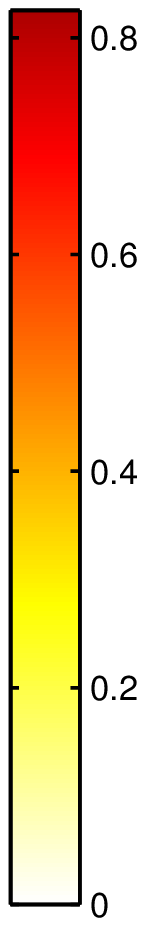} \\
	$\Sigma9$\plane114 & $\Sigma11$\plane113 & $\Sigma3$\plane112 & $\Sigma3$\plane111 & $\Sigma11$\plane332 & $\Sigma9$\plane221 &\\
	38.94$^\circ$ & 50.48$^\circ$ & 70.53$^\circ$ & 109.47$^\circ$ & 129.52$^\circ$ & 141.06$^\circ$ &
	\end{tabular}
\caption{ \label{figure3} The binding energies of HeV defect in various sites for the six \dirf110 symmetric tilt grain boundaries.  }
\end{figure}

Figure \ref{figure3} displays the spatial distribution of binding energies for HeV defects in the \dirf110 symmetric tilt grain boundary system.  Figure \ref{figure3} was arranged in the same manner as Figure \ref{figure2}, including the same extremes for the contour bar.  In this figure, the $\Sigma3$\plane112 `twin' grain boundary has a much smaller binding energy in the GB region than all other boundaries in the \dirf100 and \dirf110 STGB systems.  Interestingly, the $\Sigma3$\plane112 GB has the same disorientation angle, tilt direction, and CSL value as the $\Sigma3$\plane111 GB, but has a very different behavior in terms of the binding behavior with He defects.  Also notice that the $\Sigma3$ GBs have the lowest and highest GB energies in the present study, which can explain their difference in binding energy behavior.  For instance, in a prior study with point defects in Fe \cite{Tsc2012a}, it was found that the mean formation energies (and consequently, binding energies) for each particular grain boundary was related to the grain boundary energy.  The other nine grain boundaries showed very similar maximum binding energies and binding energy behavior within the grain boundary region, which may be more typical of general high angle grain boundaries.

\begin{figure}[ht!]
  \centering
	\begin{tabular}{ccccc}
	 \multicolumn{4}{c}{\textbf{Maximum Binding Energy}} & \footnotesize max($E^b_{HeInt}$) (eV) \\
	\cline{1-4}
	 \includegraphics[trim = 1.3in 0.0in 1.3in 0.0in, clip, height=2in,angle=0]{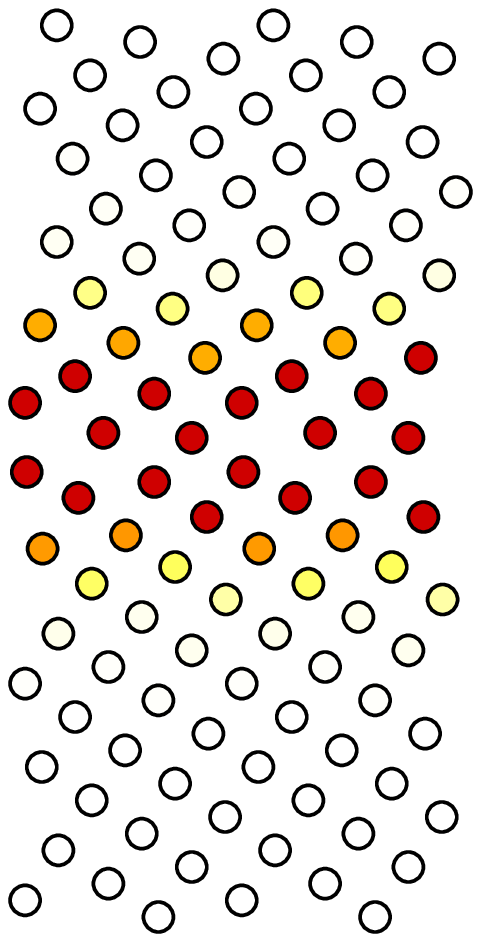} &
	 \includegraphics[trim = 1.7in 0.0in 1.7in 0.0in, clip, height=2in,angle=0]{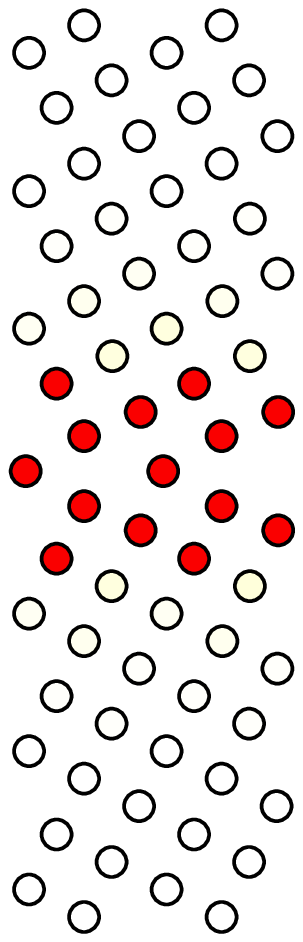} &
	 \includegraphics[trim = 1.8in 0.0in 1.8in 0.0in, clip, height=2in,angle=0]{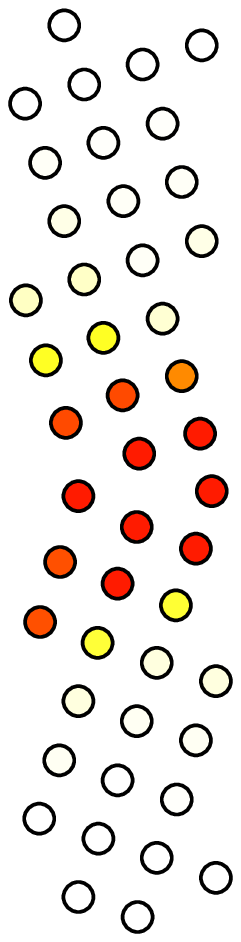} &
	 \includegraphics[trim = 1.6in 0.0in 1.6in 0.0in, clip, height=2in,angle=0]{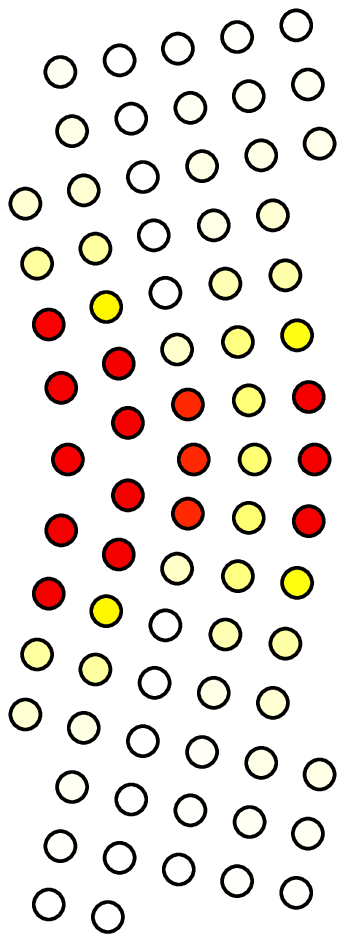} &
	 \includegraphics[trim = 4.0in 0.0in 0.0in 0.1in, clip, height=2in,angle=0]{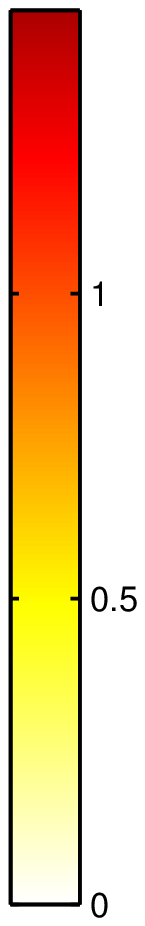} \\
	  \multicolumn{4}{c}{\textbf{Mean Binding Energy}}  & \footnotesize mean($E^b_{HeInt}$) (eV) \\
	\cline{1-4}
	 \includegraphics[trim = 1.3in 0.0in 1.3in 0.0in, clip, height=2in,angle=0]{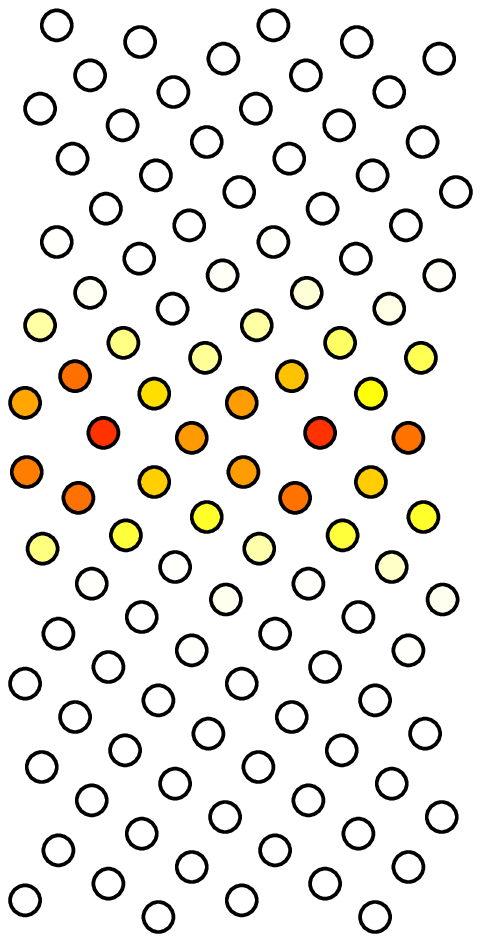} &
	 \includegraphics[trim = 1.7in 0.0in 1.7in 0.0in, clip, height=2in,angle=0]{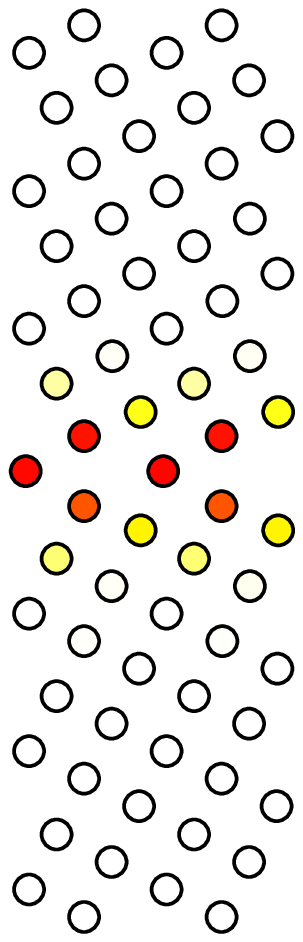} &
	 \includegraphics[trim = 1.8in 0.0in 1.8in 0.0in, clip, height=2in,angle=0]{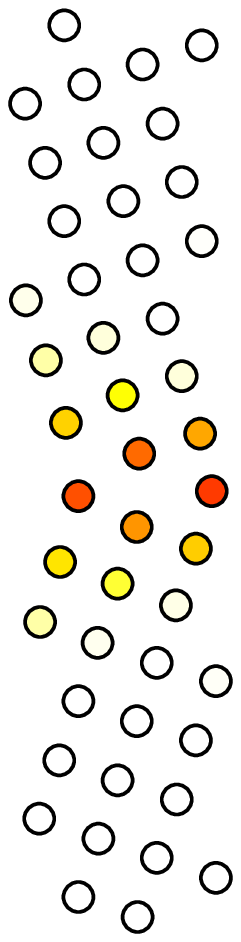} &
	 \includegraphics[trim = 1.6in 0.0in 1.6in 0.0in, clip, height=2in,angle=0]{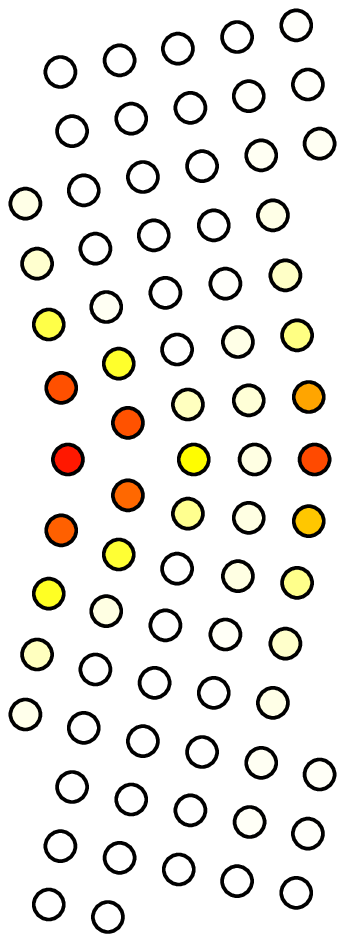} &
	 \includegraphics[trim = 4.0in 0.0in 0.0in 0.1in, clip, height=2in,angle=0]{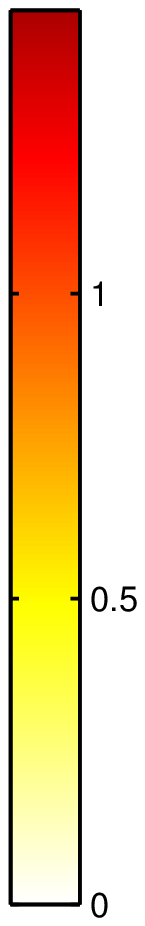} \\
	  \multicolumn{4}{c}{\textbf{Standard Deviation of the Binding Energy}}  & \footnotesize std($E^b_{HeInt}$) (eV) \\
	\cline{1-4}
	 \includegraphics[trim = 1.3in 0.0in 1.3in 0.0in, clip, height=2in,angle=0]{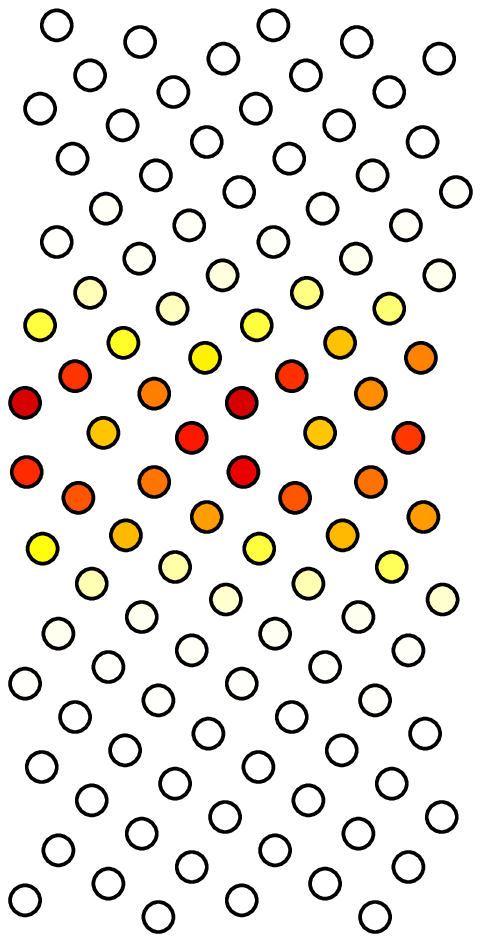} &
	 \includegraphics[trim = 1.7in 0.0in 1.7in 0.0in, clip, height=2in,angle=0]{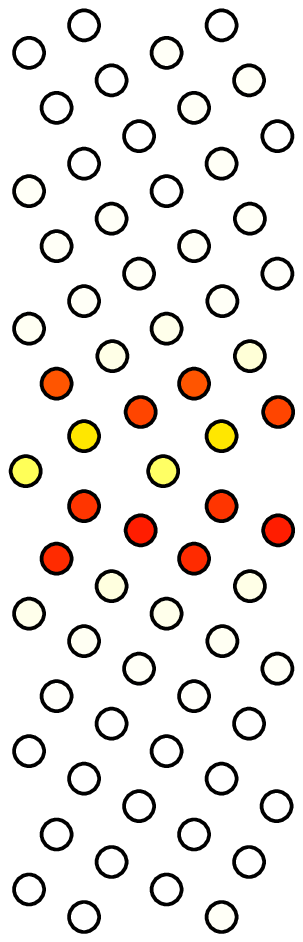} &
	 \includegraphics[trim = 1.8in 0.0in 1.8in 0.0in, clip, height=2in,angle=0]{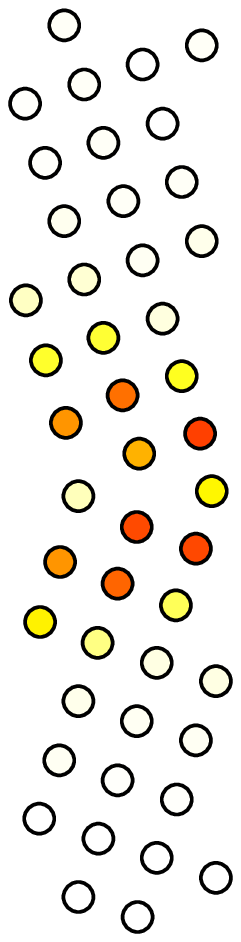} &
	 \includegraphics[trim = 1.6in 0.0in 1.6in 0.0in, clip, height=2in,angle=0]{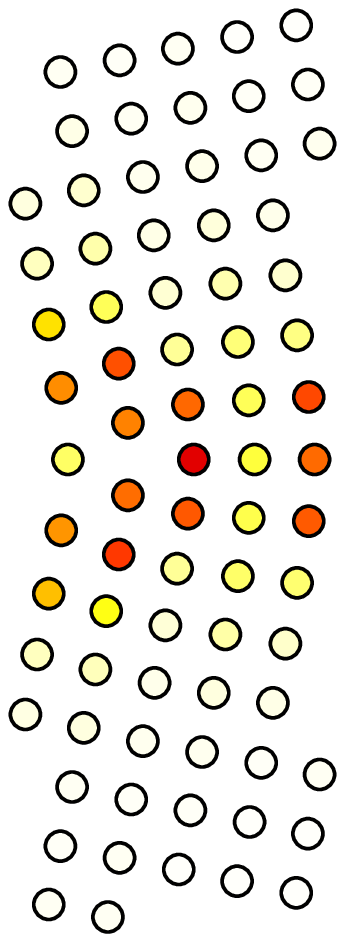} &
	 \includegraphics[trim = 4.0in 0.0in 0.0in 0.1in, clip, height=2in,angle=0]{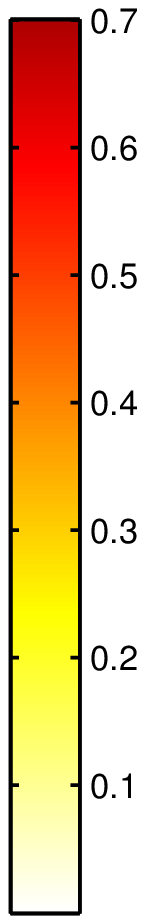} \\
	$\Sigma13$\plane510 & $\Sigma5$\plane310 & $\Sigma5$\plane210 & $\Sigma13$\plane320 & \\
	22.62$^\circ$ & 36.87$^\circ$ & 53.13$^\circ$ & 67.38$^\circ$ &
	\end{tabular}
\caption{ \label{figure4} The formation energy statistics for a He interstitial defect in various sites for the four \dirf100 symmetric tilt grain boundaries.  For each site, 20 local interstitial He locations were selected.  The formation energy statistics shown here are the minimum formation energy (top), the mean formation energy (middle), and the standard deviation of the formation energies (bottom).}
\end{figure}

The spatial distribution of binding energies for interstitial He is also of interest.  Recall that 20 different starting locations for interstitial He were run at each lattice site within 15 \AA\ of the GB center.  Hence, similar maps to Figures \ref{figure2} and \ref{figure3} can be plotted using three statistical measures of the binding energy for each site: the maximum binding energy, the mean binding energy, and the standard deviation of the binding energy.  Figure \ref{figure4} displays these three measures for the four GBs in the \dirf100 STGB system.  First, the maximum binding energy for each site indicates whether there are potential interstitial He sites surrounding each Fe atom location where the He atom would have a decreased formation energy (and large binding energy).  For instance, there is a row of atoms that bends through the grain boundary structure in the $\Sigma13$\plane320 GB that has binding energies for interstitial He similar to bulk values for all 20 locations randomly chosen around each of these atoms.  As the misorientation angle approaches 0$^\circ$ and 90$^\circ$ in the \dirf100 STGB system (i.e., lower angle GBs), there should be a larger fraction of similar areas as the distance between GB dislocations increases.  The mean binding energy indicates that although some Fe sites have high maximum binding energies, many of the 20 different He interstitial locations have much lower binding energies.  Hence, many sites that are further away from the grain boundary center in fact have much lower mean binding energies than their maximum binding energy would lead one to believe.  This measure also indicates that in some cases along the center of the GB and within certain grain boundary structural units, nearly all of the 20 random interstitial sites have high binding energies; from a statistical viewpoint, these sites may be energetically favored over time.  The standard deviation of the binding energy  can indicate the spread of the binding energies from these 20 He interstitial sites.  The low standard deviation of some sites along the grain boundary plane and within the bulk lattice indicates that most He interstitial positions have similar binding and formation energies in the area surrounding these atom positions.  The high standard deviations indicate the diffuse transition region from areas of low binding to areas of high binding areas.  For instance, in some of these locations, a random He interstitial site placed closer to the boundary center may have a much higher binding energy than a random site placed closer to the bulk crystal lattice.  While individually these three measures give information of the distribution of binding and formation energies about each particular site, the remainder of the analysis will focus on the mean binding energies of the 20 different instantiations, which is more sensitive to local variations than the maximum binding energy and is more applicable to the energetic favorability of He defects than the standard deviation would be.      

\begin{figure}[ht!]
  \centering
	\begin{tabular}{ccccc}
	 & & & & \footnotesize max($E^b_{max}$) (eV) \\
	 \includegraphics[trim = 1.6in 0.0in 1.6in 0.0in, clip, height=3in,angle=0]{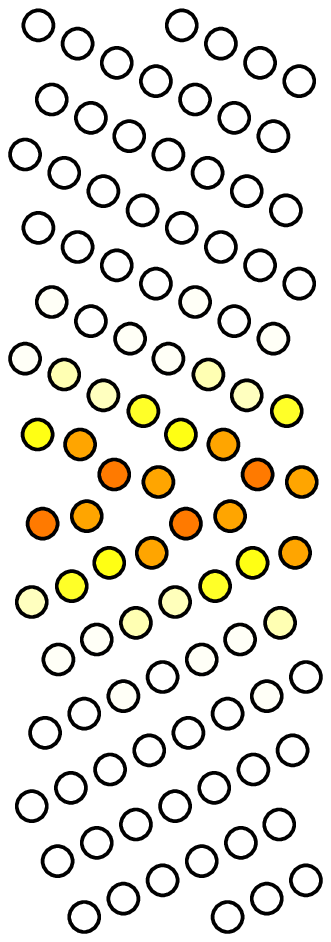} &
	 \includegraphics[trim = 1.6in 0.0in 1.6in 0.0in, clip, height=3in,angle=0]{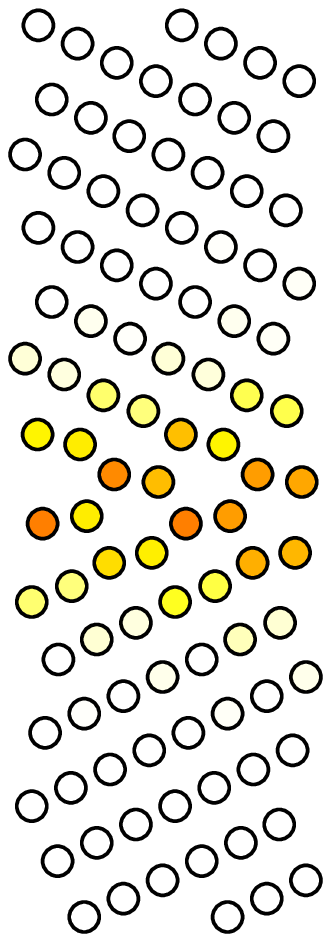} &
	 \includegraphics[trim = 1.6in 0.0in 1.6in 0.0in, clip, height=3in,angle=0]{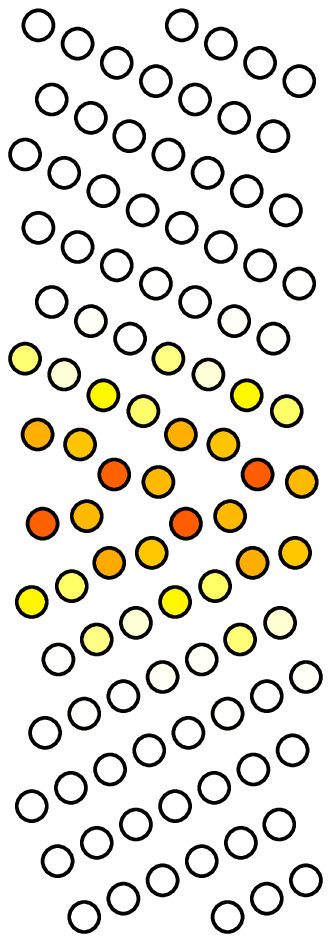} &
	 \includegraphics[trim = 1.6in 0.0in 1.6in 0.0in, clip, height=3in,angle=0]{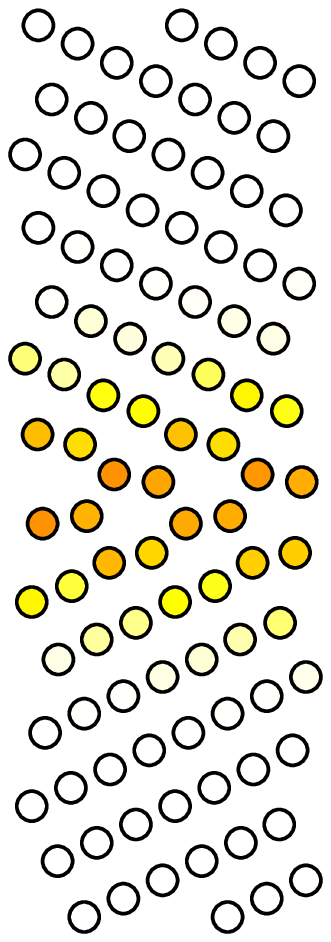} &
	 \includegraphics[trim = 4.0in 0.0in 0.25in 0.1in, clip, height=3in,angle=0]{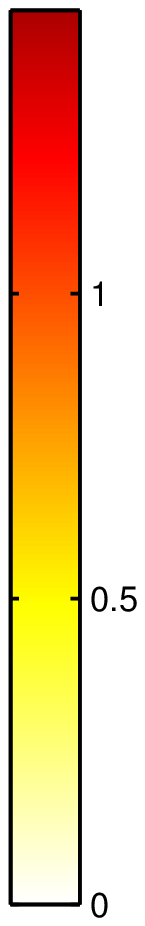} \\
	HeV & HeInt & He$_2$V & He$_2$Int & 0 eV
	\end{tabular}
\caption{ \label{figure5} The binding energies of the four different He defects in various sites for a representative high angle grain boundary ($\Sigma11$\plane332 GB).}
\end{figure}

The relative binding energy of the four He defects with respect to the grain boundary structure is also of interest.  For instance, Figure \ref{figure5} shows the spatial distribution of binding energies for the four He defects investigated in the present study for a representative high angle STGB: the $\Sigma11$\plane332 GB.  The contour bar for each defect is scaled from 0 eV to the maximum binding energy among all 10 GBs for that particular He defect.  Therefore, the represented values are a relative measure with respect to the maximum binding energy for each defect to facilitate comparison between the different He defects.  In many respects, the binding behavior is very similar between the different He defect types.  The sites with the largest binding energies are the same for all He defects.  In terms of differences, the He$_2$V and He$_{2,int}$ defects have a slightly larger interaction length scale compared to the two single He defects.  This interesting behavior may indicate that the formation and binding energies of lower order defect types may actually be an adequate predictor of the formation energies of higher order defect types, which may be important for more expensive quantum mechanics simulations.  In fact, the linear correlation coefficient $R$ was calculated for the associated formation energies for each site (the mean formation energy are used for HeInt, He$_2$V, and He$_2$Int) and all correlations were statistically significant with values of $R$ greater than 0.905 (e.g., HeV-He$_2$V: 0.9428, HeInt-He$_2$Int: 0.9606, HeV-HeInt: 0.9166, He$_2$V-He$_2$Int: 0.9484).  This finding perhaps asserts that the local environment strongly influences the He defect formation energies and that these energies are not independent of one another. 

\subsection{Influence of Local GB Structure}

\begin{figure}[b!]
  \centering
	\begin{tabular}{ccccccc}
	 \includegraphics[trim = 1.7in 0.0in 1.7in 0.0in, clip, height=2.25in,angle=0]{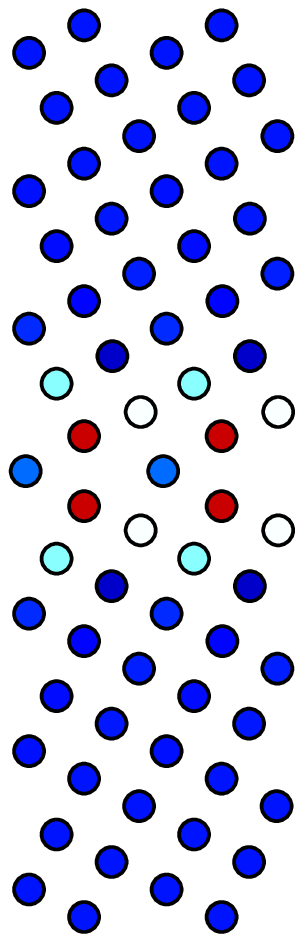} &
	 \includegraphics[trim = 1.7in 0.0in 1.7in 0.0in, clip, height=2.25in,angle=0]{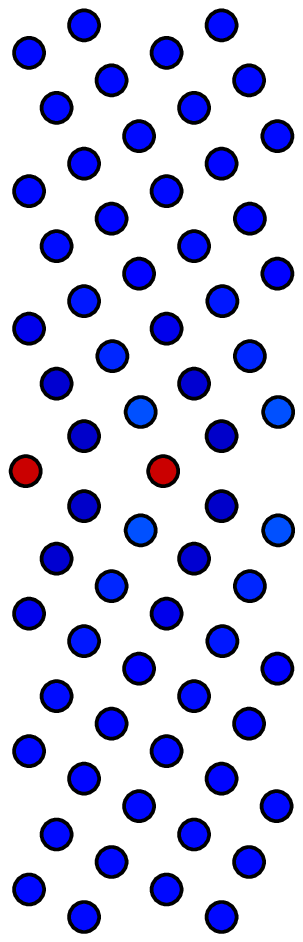} &
	 \includegraphics[trim = 1.7in 0.0in 1.7in 0.0in, clip, height=2.25in,angle=0]{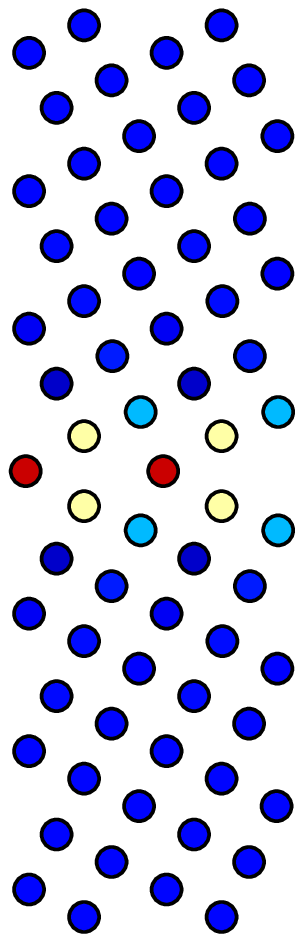} &
	 \includegraphics[trim = 1.7in 0.0in 1.7in 0.0in, clip, height=2.25in,angle=0]{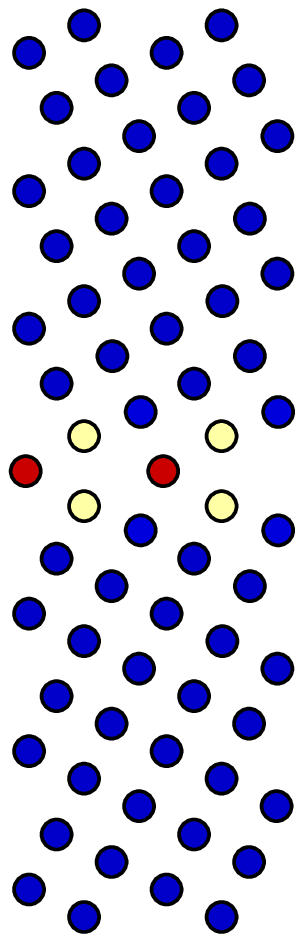} &
	 \includegraphics[trim = 1.7in 0.0in 1.7in 0.0in, clip, height=2.25in,angle=0]{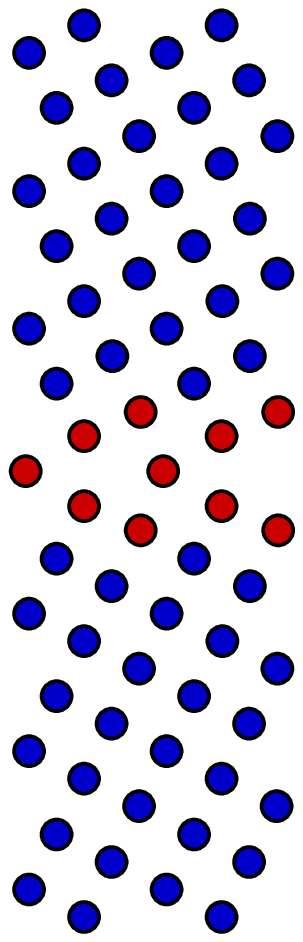} &
	 \includegraphics[trim = 1.7in 0.0in 1.7in 0.0in, clip, height=2.25in,angle=0]{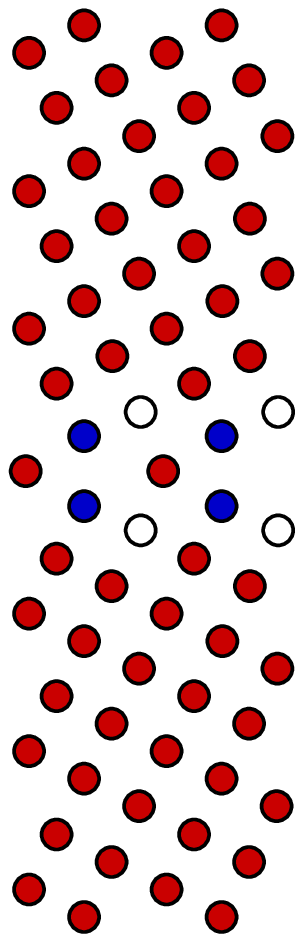} &
	 \includegraphics[trim = 4.0in 0.0in 0.0in 0.1in, clip, height=2.25in,angle=0]{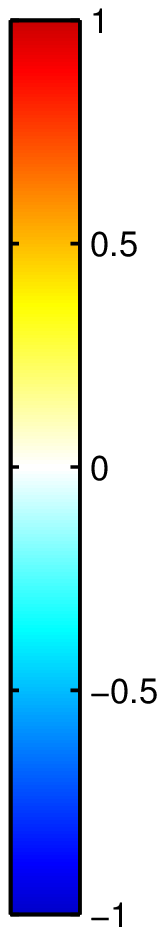} \\
	$E_{coh}$ & $\sigma_H$ & $V_{Voro}$ & CSP & CNA & CN & \\
	\end{tabular}
\caption{\label{figure6} The different local metrics performed for each grain boundary.  As an example, the $\Sigma5$\plane310 GB is shown here and the metrics are scaled from $-1$ to $+1$.  }
\end{figure}

The local environment surrounding each atom changes as due to interactions with neighboring atoms, which in turn affects the cohesive energy and other per-atom properties.  In this subsection, we will analyze several metrics used to characterize the local environment surrounding each atom and compare with the previously calculated formation and binding energies for the different He defects.  In this work, six per-atom metrics are examined: the cohesive energy $E_{coh}$, the hydrostatic stress $\sigma_H$, the Voronoi volume $V_{Voro}$, the centrosymmetry parameter CS \cite{Kel1998}, the common neighbor analysis CNA \cite{Jon1988,Hon1987,Tsu2007}, and the coordination number CN.  The per-atom virial stress components were used to calculate $\sigma_H$.  The Voronoi volume is defined from a Voronoi tesselation of the space of the Fe atoms in a clean GB.  Previous studies have shown a good correlation between the Voronoi volume and the formation energies of He defects \cite{Zha2013}.  Moreover, DFT studies have shown that increased Voronoi volumes lead to enhanced magnetic moments, higher local energies, and tensile stresses \cite{Bha2013}.  For CNA and CN, a cutoff distance of 3.5 \AA\ was used and eight nearest neighbors were used to calculate the CSP.  Over all GBs, CNA classifies the sites as 85.7\% BCC and 14.3\% unknown (866 sites total for 10 GBs).  The coordination numbers for these sites are: 1.2\% 11 CN, 2.5\% 12 CN, 4.0\% 13 CN, 91.0\% 14 CN (perfect BCC lattice), 0.6\% 15 CN, 0.7\% CN.  Hence, there are more undercoordinated sites than overcoordinated sites, as would be expected at the grain boundary (i.e., positive free volume).  Figure \ref{figure6} normalizes the six metrics and shows their evolution as a function of spatial position within the $\Sigma5$\plane310 GB.  The first four parameters in Figure \ref{figure6} can have a continuous distribution, while CNA and CN have discrete values that correlate to the crystal structure (BCC and unknown) and the local coordination number (12--14, where a coordination of 14 corresponds to 1$^{st}$ and 2$^{nd}$ nearest neighbors in a perfect BCC single crystal lattice), respectively.  There are subtle differences between the different metrics.  For example, the atom along the central plane of the GB has a coordination number of 14 (equivalent to single crystal), but has much different $\sigma_H$, $V_{Voro}$, and CSP than the bulk lattice.  Additionally, CNA identifies the atom furthest away from the boundary as unknown (similar to other GB atoms), but CSP evaluates this atom as having the same centrosymmetric environment as the bulk lattice.  Hence, these local parameters may shed light upon the ability to predict the magnitudes or trends in formation/binding energies of different He defects for different GB structures. Furthermore, the per-atom vacancy binding energies $E_b^f$ from Tschopp et al.~\cite{Tsc2011, Tsc2012a} will be compared as part of this analysis, which may be relevant since the HeV and He$_2$V defects involve a monovacancy.  The binding energies are used instead of formation energies; realize that to change the correlation coefficient $R$ from binding energies $E_b$ to formation energies $E_f$, a factor of $-1$ can be applied to all $R$ values in Tables \ref{table3} and \ref{table4}.

The per-atom metrics were tabulated for all grain boundaries and compared with each other to find which metrics are correlated.  The linear correlation coefficient $R$ is again used to compare the degree of correlation between the local metrics and the binding energies, where $R=1$ indicates a perfect positive correlation and $R=-1$ indicates a perfect negative correlation.  The results are shown in Table \ref{table3}.  The $R$ matrix is symmetric and extraneous data is removed from Table \ref{table3}.  All correlations are statistically significant ($p$-value $<0.05$) except for the two correlations with Voronoi volume.  Interesting, the binding energy of vacancies is highly positively correlated ($R=0.89$)) with the atomic cohesive energy (i.e., higher $E_{coh}$ means higher binding energy for a monovacancy).  Also, the hydrostatic stress is positively correlated ($R=0.85$) with the Voronoi volume as well, which agrees with recent DFT results \cite{Bha2013}.  Furthermore, the coordination number is negatively correlated ($R=-0.80$) with cohesive energy $E_{coh}$ (i.e., undercoordinated atoms tend to have higher $E_{coh}$). Note that for calculating the correlation coefficients from the CNA results, BCC is given a lower number than the unknown classification (i.e., unknown sites tend to have higher vacancy binding energies than BCC sites).  Since the highest $R$ value is 0.89, Table \ref{table3} indicates that the per-atom metrics examined may influence each other, but are not redundant metrics for comparing with the He defect binding energies.  

\begin{table}
\centering
\caption{\label{table3} Linear correlation coefficients $R$ for the local per-atom metrics and $E_b^f$ with each other.}
\begin{ruledtabular}
\begin{tabular}{cccccccc}
 & $E_b^f$ \tnote{b} 	& $E_{coh}$ 	& $\sigma_H$ 	& $V_{Voro}$ 	& CSP 	& CNA 	& CN \\
\hline \\ [-1.5ex]
$E_b^f$ \tnote{b} 	&   1.00 &   0.89 &  -0.45 &  -0.06\tnote{a} &   0.41 &   0.50 &  -0.66 \\ 
$E_{coh}$  &   -- &   1.00 &  -0.47 &  -0.01\tnote{a} &   0.44 &   0.63 &  -0.80 \\ 
$\sigma_H$ &  -- &  --  &   1.00 &   0.85 &   0.35 &   0.18 &   0.61 \\ 
$V_{Voro}$&  --  &  --  &  --  &   1.00 &   0.67 &   0.57 &   0.23 \\ 
CSP &   --  &   --  &   --  &   --  &   1.00 &   0.77 &  -0.24 \\ 
CNA &   --  & --  &  --  &  --  &  --  &   1.00 &  -0.44 \\ 
CN &  --  &  --  &   --  &  --  &  --  &  --  &   1.00 \\ 
\end{tabular}
        \begin{tablenotes}
		\scriptsize
            	\item [a] Not statistically significant based on a $p$-value of 0.05.
            	\item [b] The vacancy binding energy $E_b^v$ is negatively correlated ($R=-1$) with vacancy formation energy $E_f^v$.
        \end{tablenotes}
\end{ruledtabular}
\end{table}

The per-atom metrics were then compared to the binding energies of the four He defects in the current study.  The results are shown in Table \ref{table4}.  All correlations are statistically significant ($p$-value $<0.05$).  Overall the $R$ values did not deviate much over the four He defects, because of the good correlation between the binding energies of the four different He defects.  Therefore, the average $R$ value can actually give an accurate portrayal of the effect of these local environment metrics with He defects.  The CNA and CSP have the highest positive correlation ($R=0.83$ and $R=0.76$, respectively) with the binding energies of the defects.  The hydrostatic stress $\sigma_H$ and CN have the lowest correlation with the He defect binding energies.  All metrics have a positive correlation with the binding energies of the He defects except for the coordination number, where undercoordinated atoms have higher binding energies and vice versa.  Several metrics have slightly higher correlations for $E_b$ of HeV as well, indicating that as the degree of correlation decreases as the complexity of the defect increases.  That is, the binding energy of a di-He interstitial is related to several neighboring Fe atoms and may be better correlated to the per-atom metrics of multiple surrounding atoms, instead of just one.  The analysis from this subsection indicates that formation and binding energies of different He defects may be able to predicted based on computationally less expensive simulations of local metrics using conventional metamodeling techniques (polynomial regression models, radial basis functions, Kriging models, artificial neural networks, etc.).  However, this analysis also shows that the binding energies are not linearly related to one simple metric, even though the present metrics seem to be the most physically-based metrics. 

\begin{table}
\centering
\caption{\label{table4} Linear correlation coefficients $R$ for the local per-atom metrics, $E_b^f$ and the binding energies of the four He defects.}
\begin{ruledtabular}
\begin{tabular}{cccccccc}
 & $E_b^f$ 	& $E_{coh}$ 	& $\sigma_H$ 	& $V_{Voro}$ 	& CSP 	& CNA 	& CN \\
\hline \\ [-1.5ex]
$E_b^{HeV}$&  0.67 &   0.76 &   0.20 &   0.63 &   0.77 &   0.85 &  -0.46 \\ 
$E_b^{HeInt}$  &  0.57 &   0.63 &   0.28 &   0.65 &   0.77 &   0.82 &  -0.36 \\ 
$E_b^{He_2V}$ &  0.55 &   0.61 &   0.34 &   0.70 &   0.77 &   0.83 &  -0.34 \\ 
$E_b^{He_2Int}$&  0.56 &   0.63 &   0.28 &   0.63 &   0.75 &   0.80 &  -0.34 \\ 
\hline \\ [-1.5ex]
Average &  0.59 &   0.66 &   0.27 &   0.65 &   0.76 &   0.83 &  -0.37 \\ 
\end{tabular}
\end{ruledtabular}
\end{table}

\begin{figure}[ht!]
  \centering
        \begin{subfigure}[b]{0.475\textwidth}
  \centering
	 \includegraphics[width = 0.9\textwidth,angle=0]{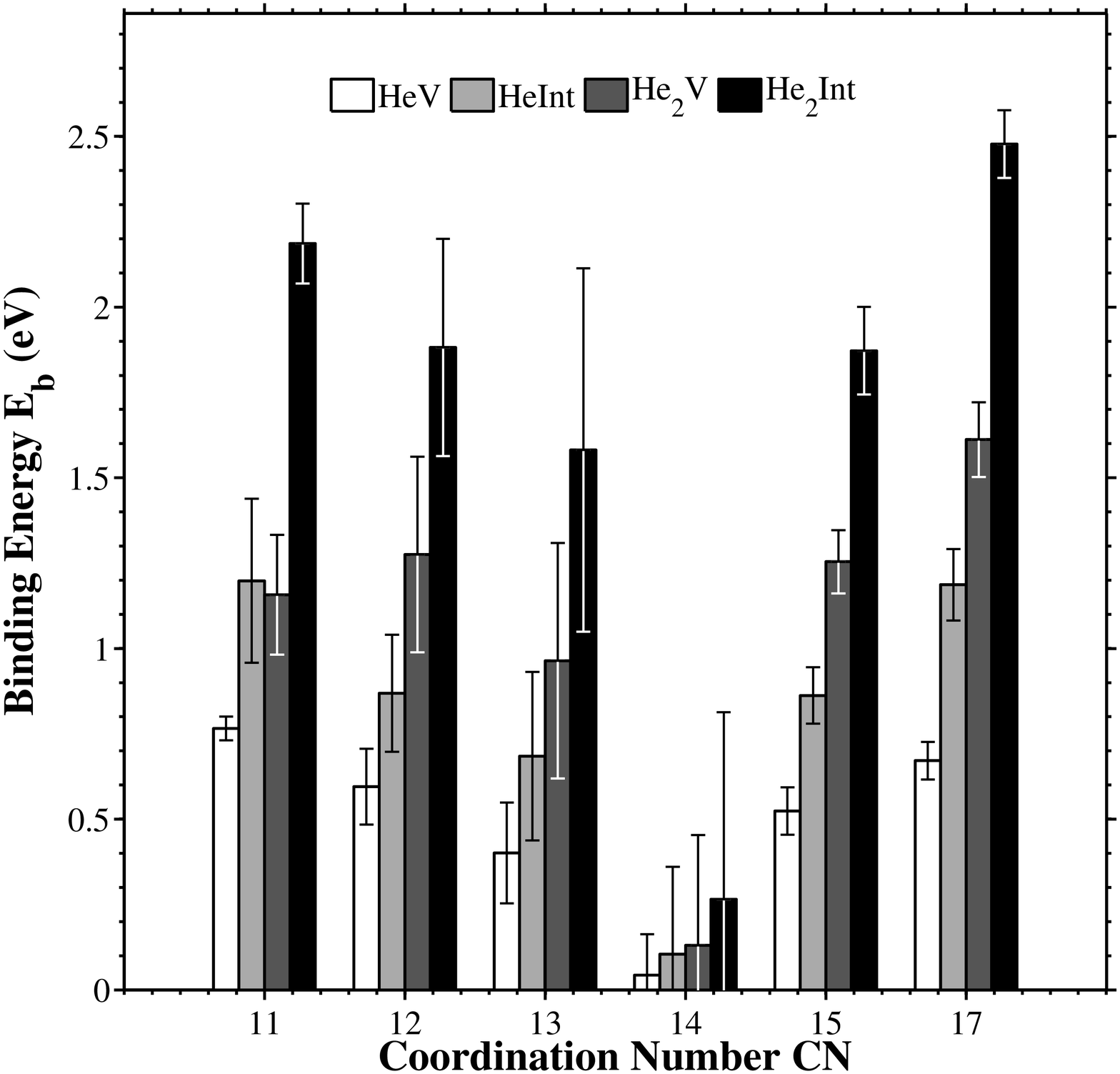}
                \caption{Coordination Number}
                \label{fig19a}
        \end{subfigure}%
        \begin{subfigure}[b]{0.475\textwidth}
  \centering
	 \includegraphics[width = 0.9\textwidth,angle=0]{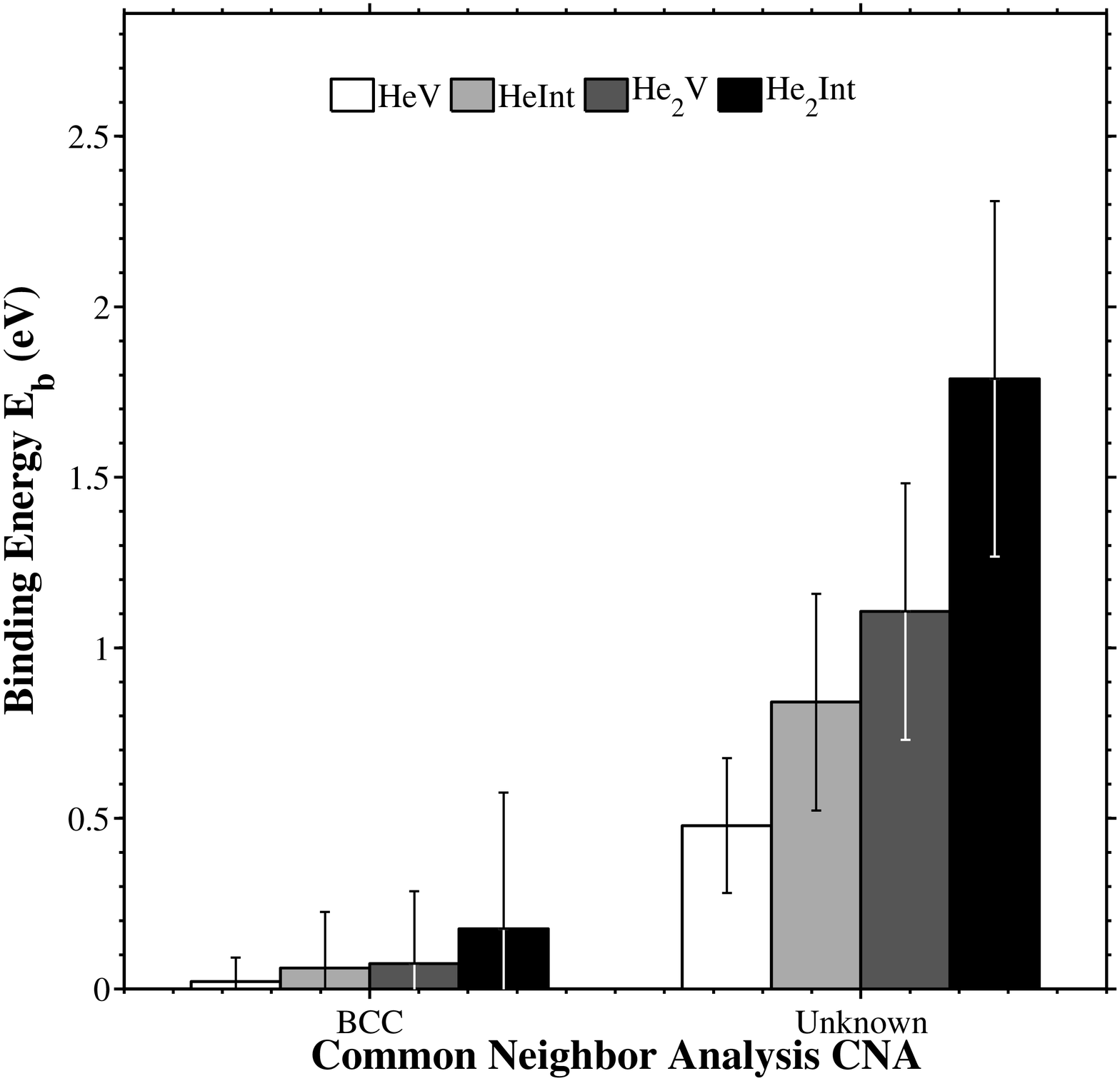}
                \caption{Common Neighbor Analysis}
                \label{fig19b}
        \end{subfigure}
\caption{\label{figure19}The influence of (a) coordination number and (b) common neighbor analysis classification on the binding energy of the four He defects.  The bar height corresponds to the mean binding energy and the error bars correspond to one standard deviation of the binding energy distribution.  The percentage of atoms that fall into these classifications are as follows: CN -- 1.2\%,    2.5\%,    4.0\%,   91.0\% (14),    0.6\%,   0.7\% (left to right), and CNA -- 85.7\% (BCC), 14.3\% (unknown).}
\end{figure}

Figure \ref{figure19} further examines the correlation with the coordination number and common neighbor analysis for the binding energies of the four He defects.  In both plots, the error bar represents one standard deviation of the binding energy distribution for each particular He defect.  For coordination number, the perfect BCC coordination number of 14 has the lowest mean binding energies for the four defects.  The standard deviation is large for perfect BCC coordination because most atoms have no binding energy in the bulk lattice, but there are a few atoms within the grain boundary region that have a much higher binding energy.  This plot also shows that both under- and over-coordinated atoms have a higher binding energy.  Moreover, the degree of undercoordination increases the binding energy for most He defect types.  The common neighbor analysis also shows a definitive difference between those atoms identified as BCC (mainly bulk lattice atoms) and those identified as unknown (mainly GB atoms, see Figure \ref{figure6}).  Another classification per-atom metric is the Ackland analysis \cite{Ack2006}, which can be used to classify atoms into five categories based on their local environment: unknown, BCC, FCC, HCP, and icosahedral.  This method attempts to differentiate the crystal structure in dynamic simulations by distinguishing characteristics between the various crystal structures.  Similar to the common neighbor analysis, the Ackland analysis also shows a distinct separation in binding energies between BCC and other non-BCC crystal structures for all He defects.  While there are some minor differences in the mean and standard deviations of the non-BCC crystal structures, the sample size is perhaps too small to statistically confirm that these non-BCC structures are different (e.g., the icosahedral cluster classification consists of 4 atoms, perhaps explaining the small standard deviations). 

\begin{figure}[ht!]
  \centering
	 \includegraphics[width = 0.8\textwidth,angle=0]{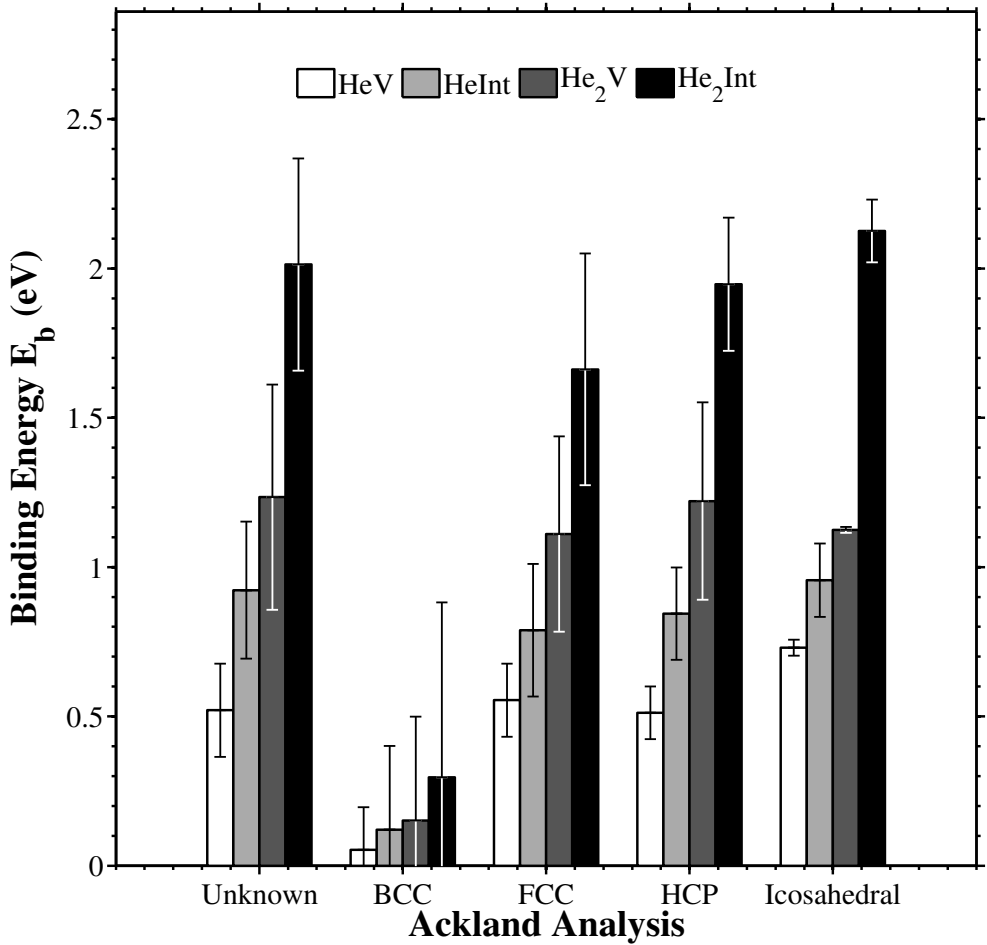}
\caption{\label{figure20}The influence of the Ackland analysis classification on the binding energy of the four He defects.  As with Figure \ref{figure19}, the bar height corresponds to the mean binding energy and the error bars correspond to one standard deviation of the binding energy distribution.    The percentage of atoms that fall into these classifications are as follows: 3.3\%,   93.2\% (BCC),    1.8\%,    1.2\%,    0.5\% (left to right).}
\end{figure}

\subsection{Statistical GB Description}

\begin{figure}[b!]
  \centering
	\includegraphics[width=0.8\columnwidth,angle=0]{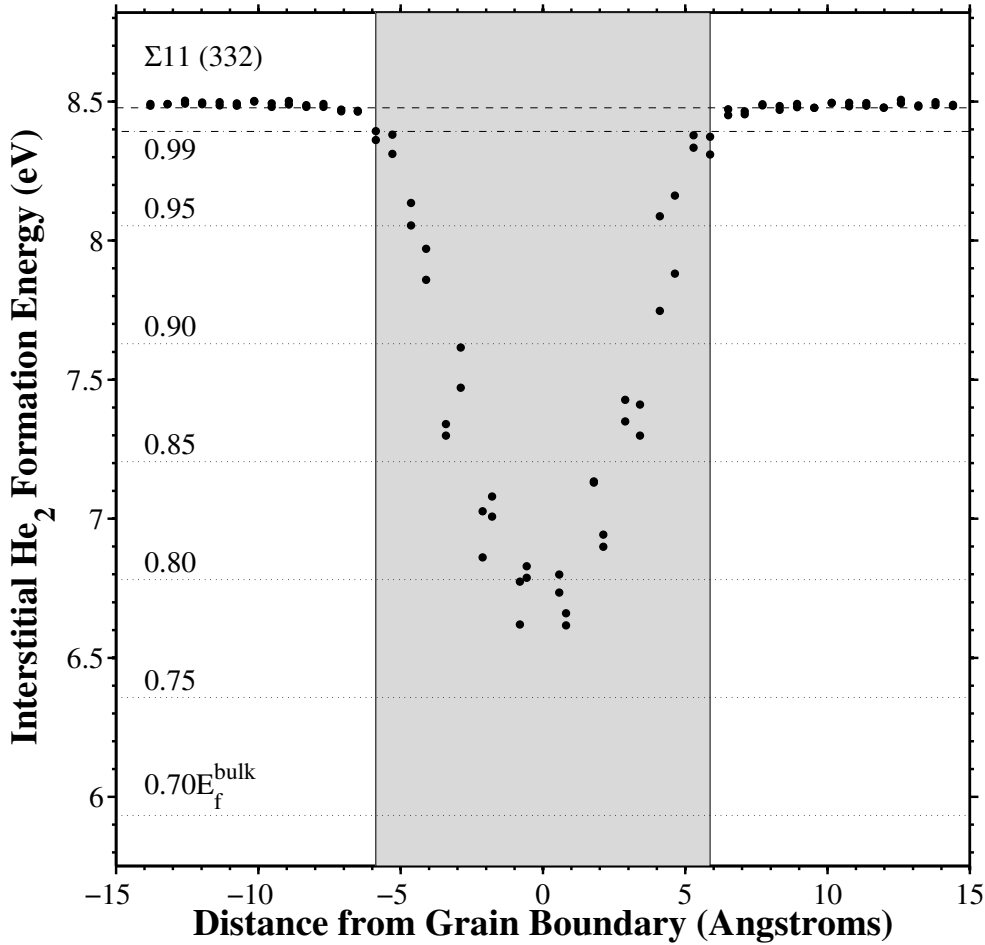} \\
\caption{ \label{figure7} The formation energies of interstitial He$_2$ defect in various sites plotted against the distance from the grain boundary for the (a) $\Sigma3$\plane112 and (b) $\Sigma11$\plane332 GBs.  }
\end{figure}

The formation energies of the various He defects can be plotted against the distance from the grain boundary to quantify the evolution of the formation energies (and binding energies) near the GB and to quantify the length scale associated with the He defects.  Figure \ref{figure7} is an example of one such plot for the interstitial He$_2$ defect at various sites at the $\Sigma11$\plane332 GB, which has very similar behavior to most of the other GBs studied herein (aside from the $\Sigma3$\plane112 GB, to be discussed later).  In this plot, the formation energies of each He defect was first calculated for each site.  In the case of those He defects that required multiple locations per Fe atomic site, the mean formation energy was used, as in Figure \ref{figure7}.  Next, a grain boundary region was defined to compare the different He defect types and the different GBs examined in the current work. This grain boundary region was subsequently used to quantify three parameters: the segregation length scale $l_{GB}$, the mean binding energy $E_b^{mean}$, and the maximum binding energy $E_b^{max}$.  First, the difference between the mean formation energy at $>$10 \AA\ and the formation energy calculated in a 2000 atom single crystal unit cell was calculated for each boundary and any bias detected was subsequently removed.   Prior simulations to test for convergence of formation energies as a function of  simulation cell size show that this bias was associated with the simulation cell size.  The simulation cell sizes given in Table \ref{table1} produced a bias on the order of $0.01E_f^{bulk}$.  The GB-affected region is now defined by identifying all formation energies that deviate more than $0.01E_f^{bulk}$ from the bulk $E_f^{bulk}$ (i.e., $0.99E_f^{bulk}$ in Figure \ref{figure7}); minimum length that encompasses all these GB sites is $l_{GB}$.  The GB-affected region calculated using this criterion is shaded light gray in Fig.~\ref{figure7}.  Next, the formation energies were converted to binding energies for each He defect/GB combination to compare the energy gained by each defect segregating to the boundary as opposed to the bulk lattice.  Both the mean and maximum binding energies ($E_b^{mean}$ and $E_b^{max}$, respectively) for this region are then calculated.  To illustrate the percent difference from the bulk formation energy, increments of $0.05E_f^{bulk}$, or 5\% of the bulk formation energy, are indicated by dotted lines in Figure \ref{figure7}.  For instance, in the $\Sigma11$\plane332 GB, the maximum binding energy is over 20\% of the bulk formation energy and lies towards the center of the grain boundary region.  This technique for identifying three parameters for He segregation was subsequently applied to all 10 GBs for all defect types.

\begin{table}
\centering
\caption{\label{table5} Binding behavior and length scales of various He defects to the ten grain boundaries in the present work}
\begin{ruledtabular}
\begin{tabular}{cccccccc}
\multirow{3}*{He Atoms} & \multirow{3}*{GB Type}& \multicolumn{3}{c}{Substitutional He} & \multicolumn{3}{c}{Interstitial He} \\
\cline{3-8} \\ [-1.5ex]
&  & $l_{GB}$ & $E_{b}^{mean}$ & $E_{b}^{max}$ & $l_{GB}$ & $E_{b}^{mean}$ & $E_{b}^{max}$ \\
&  & (\AA) & (eV) & (eV) & (\AA) & (eV) & (eV) \\
\hline \\ [-1.5ex]
\multirow{10}*{1} &      $\Sigma3$\plane111 &   7.0 &  0.41 &  0.79 &  8.55 &  1.00 &  1.47 \\ 
&      $\Sigma3$\plane112 &   3.5 &  0.07 &  0.13 &  3.54 &  0.11 &  0.14 \\ 
&      $\Sigma5$\plane210 &   8.0 &  0.37 &  0.75 &  9.27 &  0.52 &  1.06 \\ 
&      $\Sigma5$\plane310 &   7.5 &  0.33 &  0.68 &  5.76 &  0.68 &  1.21 \\ 
&      $\Sigma9$\plane221 &   8.3 &  0.35 &  0.79 & 10.62 &  0.55 &  1.34 \\ 
&      $\Sigma9$\plane114 &   6.4 &  0.45 &  0.71 & 10.43 &  0.56 &  1.08 \\ 
&     $\Sigma11$\plane113 &   8.0 &  0.31 &  0.75 &  9.68 &  0.50 &  1.22 \\ 
&     $\Sigma11$\plane332 &   8.2 &  0.29 &  0.49 &  9.27 &  0.44 &  0.86 \\ 
&     $\Sigma13$\plane510 &   7.5 &  0.35 &  0.79 &  9.23 &  0.48 &  1.08 \\ 
&     $\Sigma13$\plane320 &  13.1 &  0.19 &  0.83 & 17.17 &  0.25 &  1.16 \\ 
\hline \\ [-1.5ex]
\multirow{10}*{2} &      $\Sigma3$\plane111 &   8.5 &  1.05 &  1.49 & 13.52 &  1.27 &  2.42 \\ 
&      $\Sigma3$\plane112 &   3.5 &  0.15 &  0.17 &  4.68 &  0.17 &  0.25 \\ 
&      $\Sigma5$\plane210 &   9.3 &  0.76 &  1.69 & 11.17 &  1.17 &  2.17 \\ 
&      $\Sigma5$\plane310 &   5.8 &  0.88 &  1.35 &  8.44 &  1.24 &  2.17 \\ 
&      $\Sigma9$\plane221 &  10.6 &  0.71 &  1.56 & 11.13 &  1.20 &  2.49 \\ 
&      $\Sigma9$\plane114 &   9.1 &  0.78 &  1.52 & 11.80 &  1.17 &  2.38 \\ 
&     $\Sigma11$\plane113 &  10.2 &  0.61 &  1.75 & 10.63 &  1.13 &  2.59 \\ 
&     $\Sigma11$\plane332 &   9.3 &  0.66 &  1.22 & 11.75 &  0.99 &  1.87 \\ 
&     $\Sigma13$\plane510 &   8.7 &  0.66 &  1.86 & 10.32 &  0.93 &  2.46 \\ 
&     $\Sigma13$\plane320 &  16.8 &  0.37 &  1.78 & 17.17 &  0.61 &  2.47 \\ 
\end{tabular}
\end{ruledtabular}
\end{table}

\begin{figure}[t!]
  \centering
	\begin{tabular}{cc}
	 \includegraphics[width=0.45\columnwidth,angle=0]{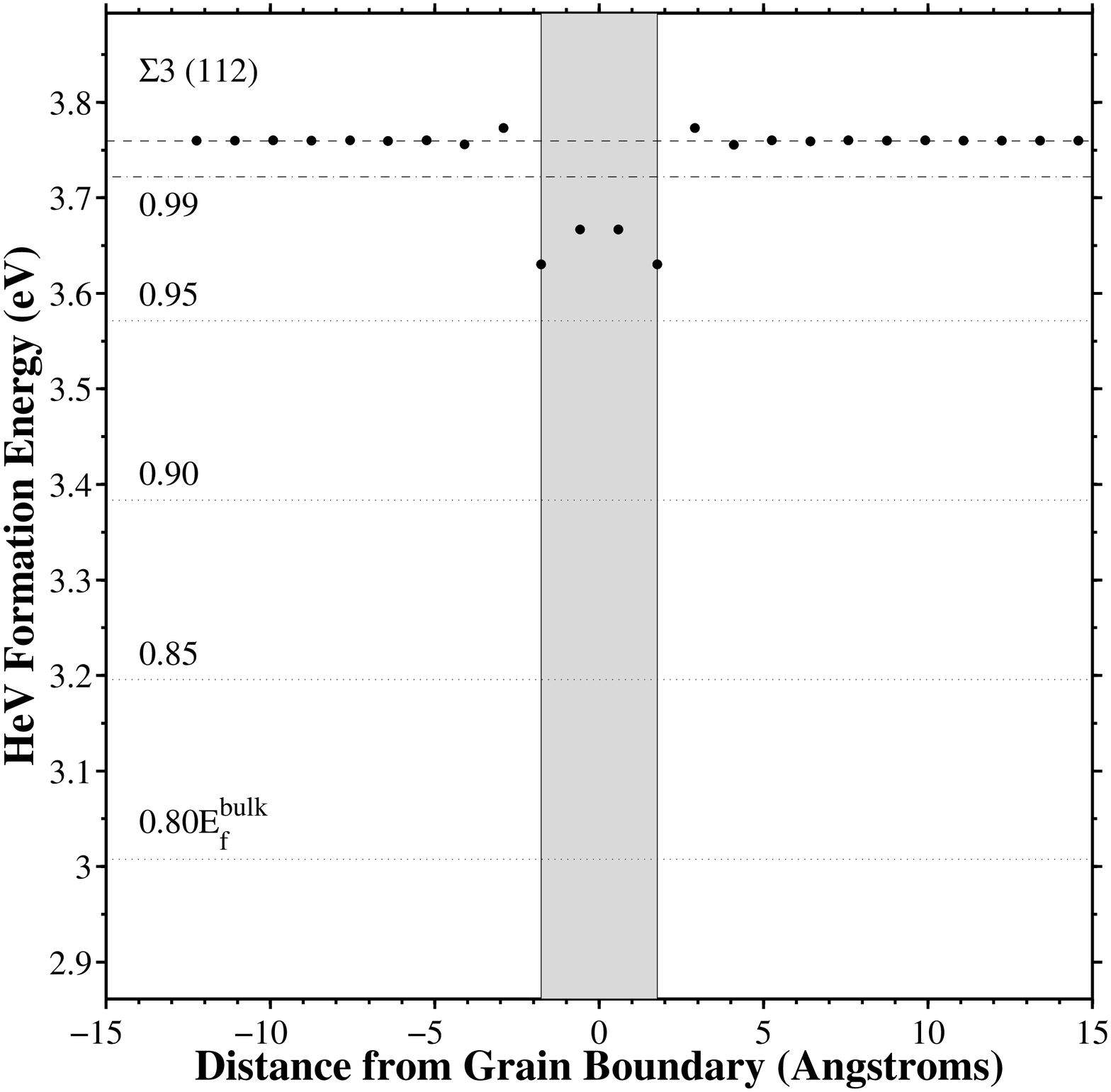} &
	 \includegraphics[width=0.45\columnwidth,angle=0]{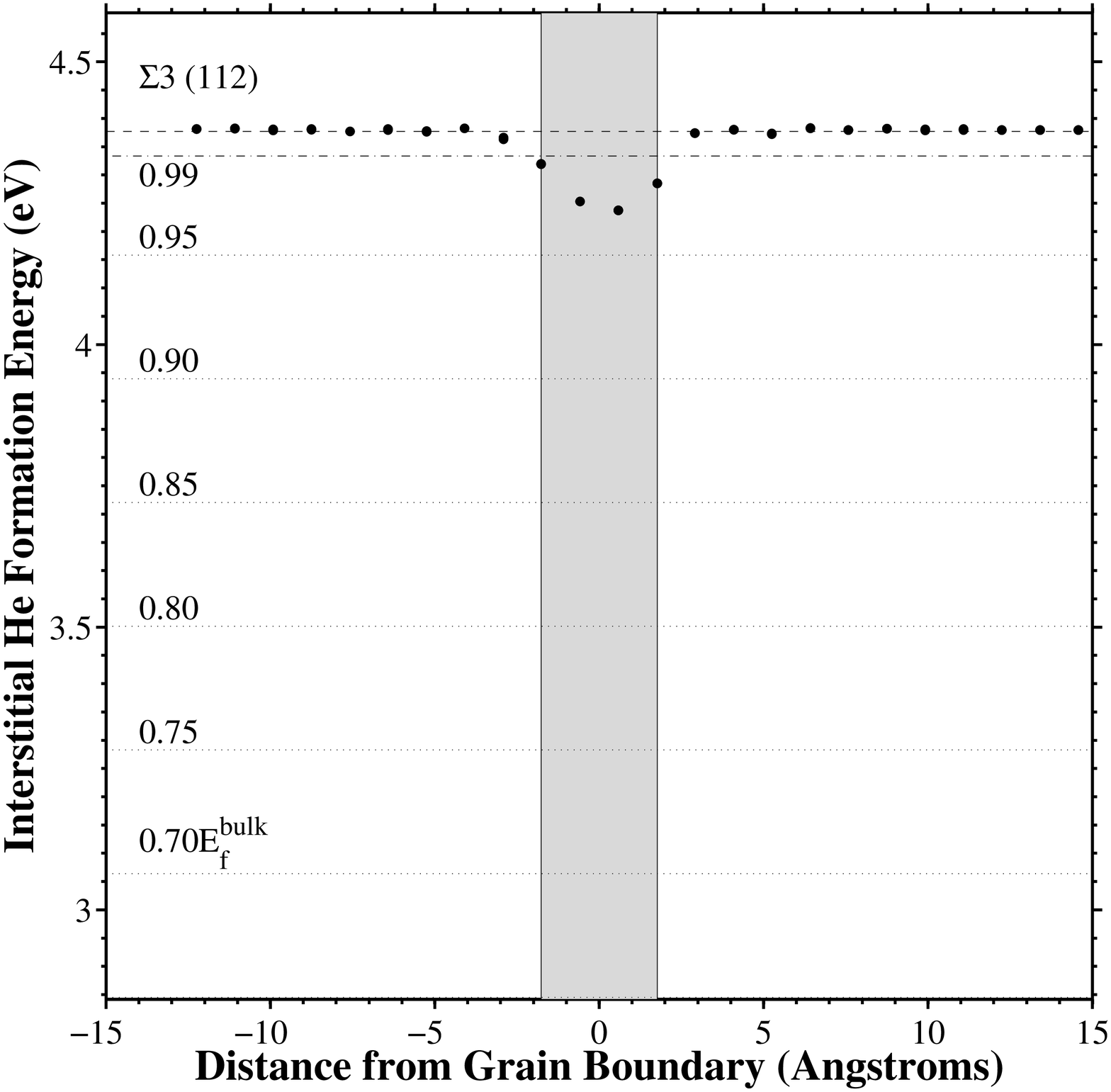} \\
	(a) HeV & (b) HeInt \\
	 \includegraphics[width=0.45\columnwidth,angle=0]{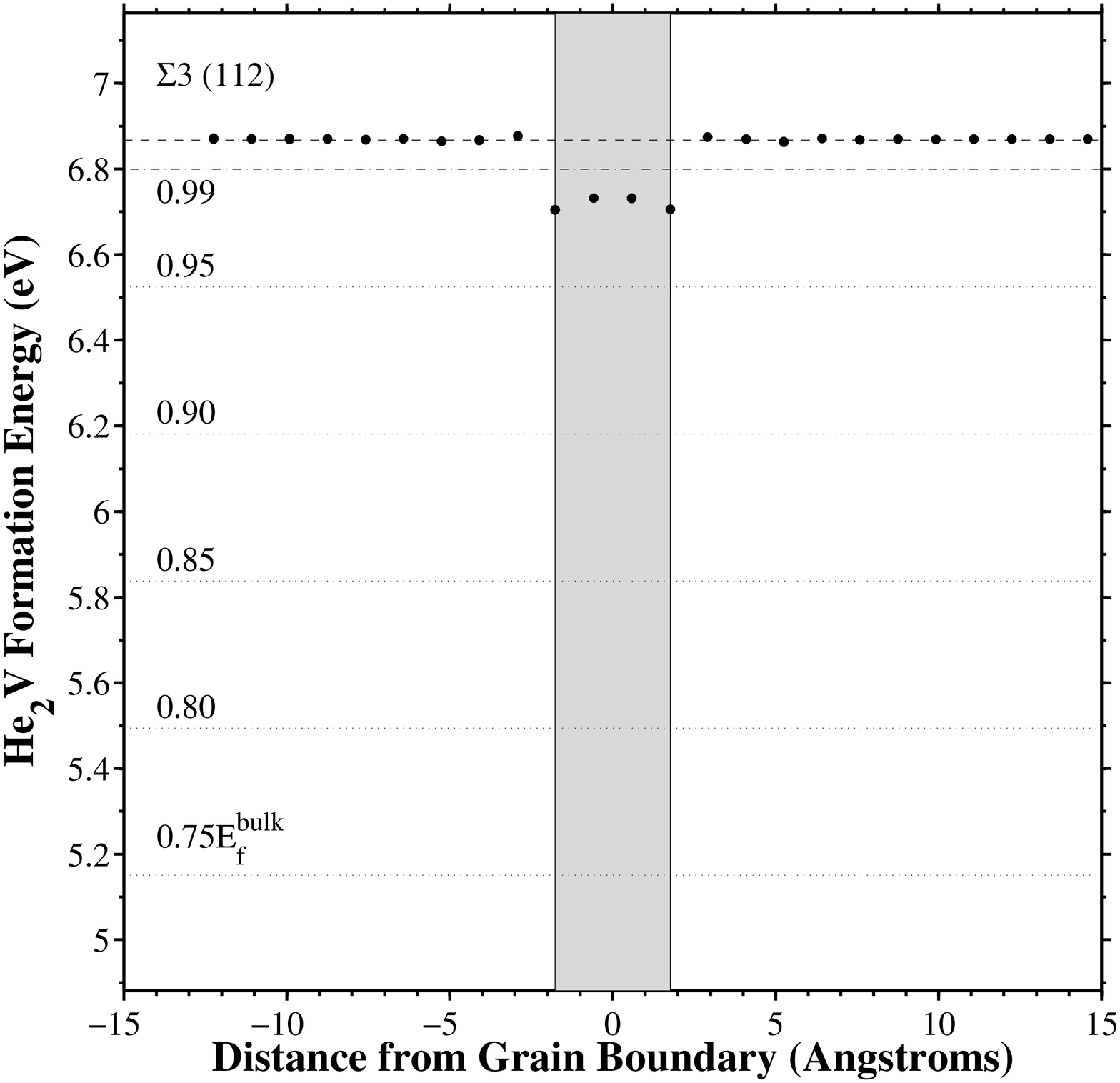} &
	 \includegraphics[width=0.45\columnwidth,angle=0]{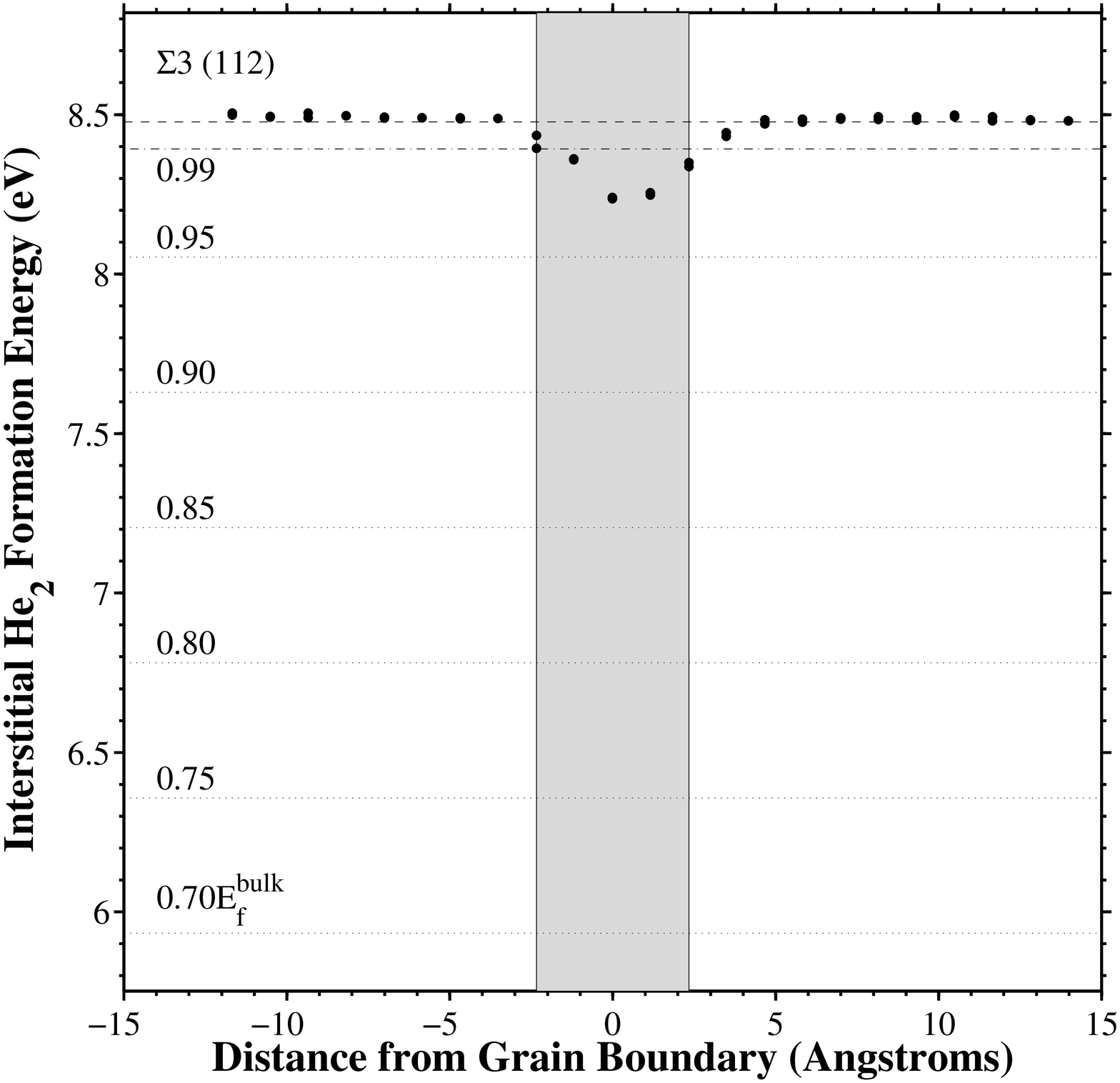} \\
	(c) He$_2$V & (d) He$_2$Int 
\end{tabular}
\caption{ \label{figure8} The formation energies of interstitial He$_2$ defect in various sites plotted against the distance from the grain boundary for the (a) $\Sigma3$\plane112 and (b) $\Sigma11$\plane332 GBs.  }
\end{figure}

The binding behavior of all four He defects to the grain boundaries examined in the current study is tabulated in Table \ref{table5}.  For the He$_{int}$, He$_2$V, and He$_{2,int}$ defects, the mean binding energy is first calculated for each site from the 20 random instantiations ($E_b^\alpha$).  Then, an overall mean and maximum binding energy ($E_b^{mean}$ and $E_b^{max}$, respectively) for the GB-affected region is attained from the set of mean binding energies.  From Table \ref{table5}, it is immediately apparent that the $\Sigma3$\plane112 twin boundary has a smaller length scale and smaller binding energies than the other boundaries.  As observed in Table \ref{table1}, this boundary has both the lowest energy and lowest free volume, which supports that these macroscale GB parameters may indicate lower binding energies with He defects (e.g., as suggested by Kurtz et al.~\cite{Kur2004}).  The other boundaries have very similar length scales (typically between 8 \AA\ to 12 \AA) and binding energies (sensitive to He defect type), with a few instances where one boundary has an interaction length scale or binding energies different from the rest (e.g., the $\Sigma13$\plane320 GB consistently has the largest GB-affected region).  In general, the GB-affected length scale is larger for the two He defect types, which may be due to a greater probability of at least one He atom being randomly placed closer to the grain boundary plane (i.e., the mean formation and binding energies would be more affected).  Moreover, the binding energies for interstitial atom(s) to the GB are higher than those for He atom(s) in a monovacancy to the GB.  Also, the binding energies for He defects with two He atoms are higher than those with one He atom.  For instance, interstitial He is more strongly bound ($E_b^{max}=$ 0.86--1.47 eV, excluding $\Sigma3$\plane112 GB) to the GB than substitutional He ($E_b^{max}=$ 0.49--0.83 eV, excluding $\Sigma3$\plane112 GB), in agreement with previous studies \cite{Kur2004} (interstitial He, $E_b^{max}=$ 0.5--2.7 eV; substitutional He, $E_b^{max}=$ 0.2--0.8 eV).  This trend holds for two He atoms as well.  Since the $\Sigma3$\plane112 GB displays a much smaller binding energy and different behavior, Figure \ref{figure8} shows the same plot as Figure \ref{figure7} for the four defects in this boundary.  The behavior is once again very similar between the different defect types, with a maximum binding energy that is approximately 2--3\% of the bulk formation energy for each He defect type.  Furthermore, the length scale associated with this boundary is much smaller than the other boundaries.

\begin{table}
\centering
\caption{\label{table6} Comparison of Fe--He interatomic potential \cite{Gao2011} and DFT \cite{Zha2010i,Zha2013z} formation and binding energies of various He defects in the $\Sigma5$\plane310 GB}
\begin{threeparttable}[b]
\begin{ruledtabular}
\begin{tabular}{cccccccc}
\multirow{3}*{He Atoms} & \multirow{3}*{Model}& \multicolumn{3}{c}{Substitutional He} & \multicolumn{3}{c}{Interstitial He} \\
\cline{3-8} \\ [-1.5ex]
&  & $E_f^{min}$ & $E_{b}^{mean}$ & $E_{b}^{max}$ & $E_f^{min}$ & $E_{b}^{mean}$ & $E_{b}^{max}$ \\
&  & (eV) & (eV) & (eV) & (eV) & (eV) & (eV) \\
\hline \\ [-1.5ex]
\multirow{3}*{1} &     MD \cite{Gao2011} &   3.08 &  0.33 &  0.68 &  3.17 &  0.68 &  1.21 \\ 
 &      DFT \cite{Zha2010i} &   2.93 &  0.68\tnote{a} &  1.20 &  2.98 &  1.14\tnote{b} &  1.43 \\ 
 & Difference & +5.1\%   & -51.5\%\tnote{e}  & -43.3\%  &  +6.4\% & -40.4\%\tnote{e}  & -15.4\%  \\ 
\hline \\ [-1.5ex]
\multirow{3}*{2} &      MD \cite{Gao2011} &   5.52 &  0.88 &  1.35 &  6.32 &  1.24 &  2.17 \\ 
&      DFT \cite{Zha2013z} &   5.08 &  1.67\tnote{c} &  1.88 &  5.62 &  2.78\tnote{d} &  2.78 \\ 
 & Difference & +8.7\%  & -47.3\%\tnote{e}  & -28.2\%  & +12.5\%  & -55.4\%\tnote{e}  & -21.9\%  \\ 
\end{tabular}
\end{ruledtabular}
        \begin{tablenotes}
		\scriptsize
            	\item [a] Binding energies of 0.64, 1.20, 0.58, and 0.28 eV for 4 different He sites.
            	\item [b] Binding energies of 1.29, 0.71, and 1.43 eV for 3 different He sites.
            	\item [c] Binding energies of 1.88 and 1.45 eV for 2 different He sites.
            	\item [d] Only one binding energy of 2.78 eV.
            	\item [e] MD calculations used larger GB region than DFT.  See text for explanation.
        \end{tablenotes}
\end{threeparttable}
\end{table}

The present calculations can also be compared to recent DFT formation/binding energies for the $\Sigma5$\plane310 GB \cite{Zha2010i,Zha2013z}.  For instance, Table \ref{table6} compares the minimum formation energy, mean binding energy, and maximum binding energy calculated herein with those from Zhang et al.~\cite{Zha2010i,Zha2013z}.  The minimum formation energies for the Fe--He interatomic potential \cite{Gao2011} were calculated using the maximum binding energies $E_b^{max}$ from Table \ref{table5} and the formation energies from Table \ref{table2} (i.e., $E_f^{min} = E_f^{bulk} - E_b^{max}$).  The trend of the data with respect to the ordering of formation energies for different He defect clusters agrees well between the Fe--He interatomic potential and DFT.  The minimum formation energies for the $\Sigma5$\plane310 GB are within $12.5$\% of each other, with a larger discrepancy in the binding energies.  The largest $E_b^{max}$ deviation is for the HeV cluster ($-43$\%), with the other He defects falling within 30\% of DFT values.  The interstitial He defect energies tend to have better agreement with DFT in terms of maximum binding energies.  Also, note that the difference is not random; formation energies are consistently higher for the Fe--He potential and binding energies are consistently lower.  While the mean values $E_b^{mean}$ have the largest difference from DFT values, note that the DFT calculations have \textit{at most} four different sites for each He defect, which are primarily located along the center plane of the grain boundary.  Hence, the mean binding energy calculated from these values is biased towards higher values, as the $E_b^{mean}$ values from the Fe--He potential in Table \ref{table5} take into account locations over the defined length scale $l_{GB}$; recall that the binding energy decreases with increasing distance from the GB center.  Hence, the present Fe--He interatomic potential \cite{Gao2011} qualitatively agrees with previous DFT calculations \cite{Zha2010i,Zha2013z} and both formation/binding energies agree with DFT within the calculated differences. 

\begin{figure}[t!]
  \centering
	\begin{tabular}{cc}
	 \includegraphics[width=0.45\columnwidth,angle=0]{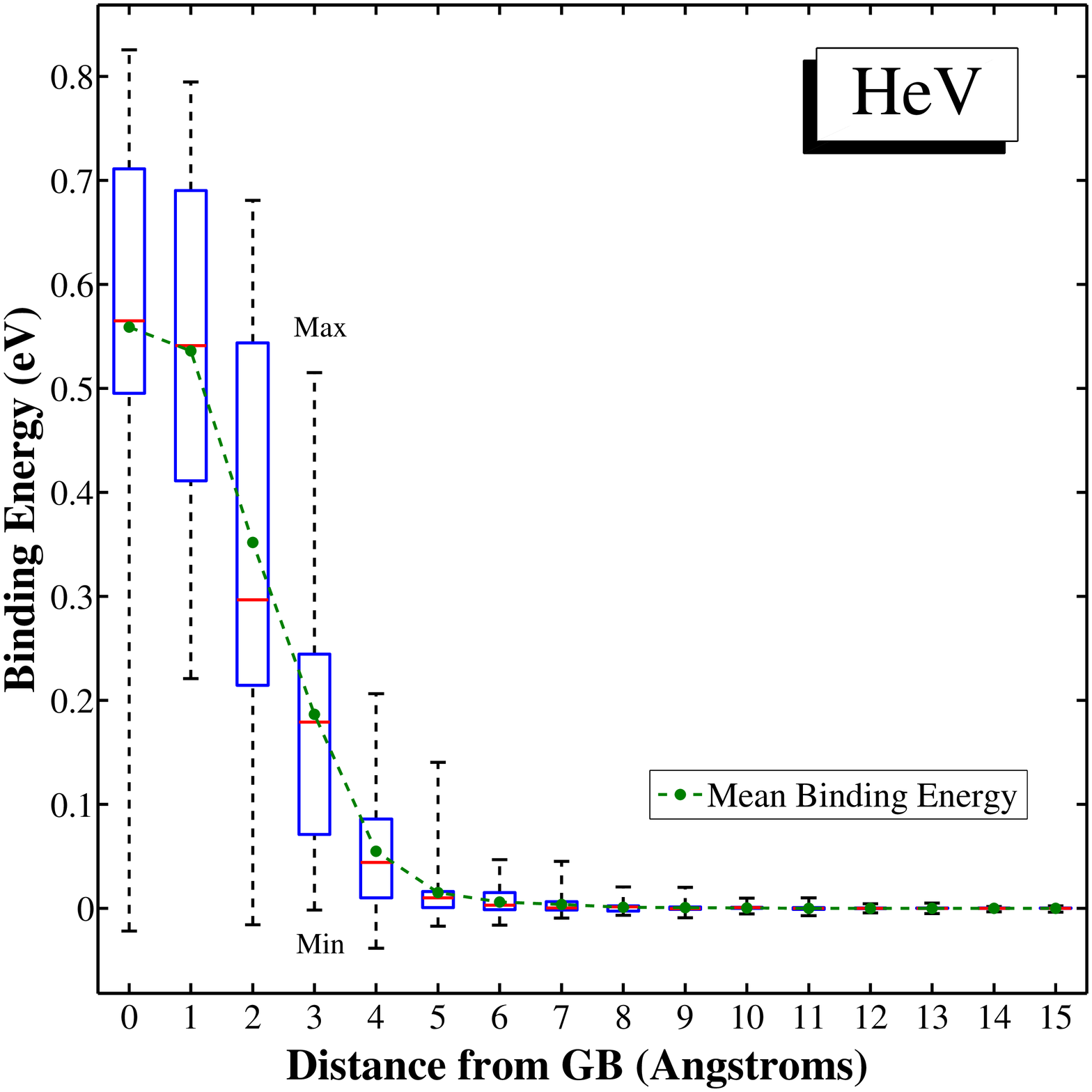} &
	 \includegraphics[width=0.45\columnwidth,angle=0]{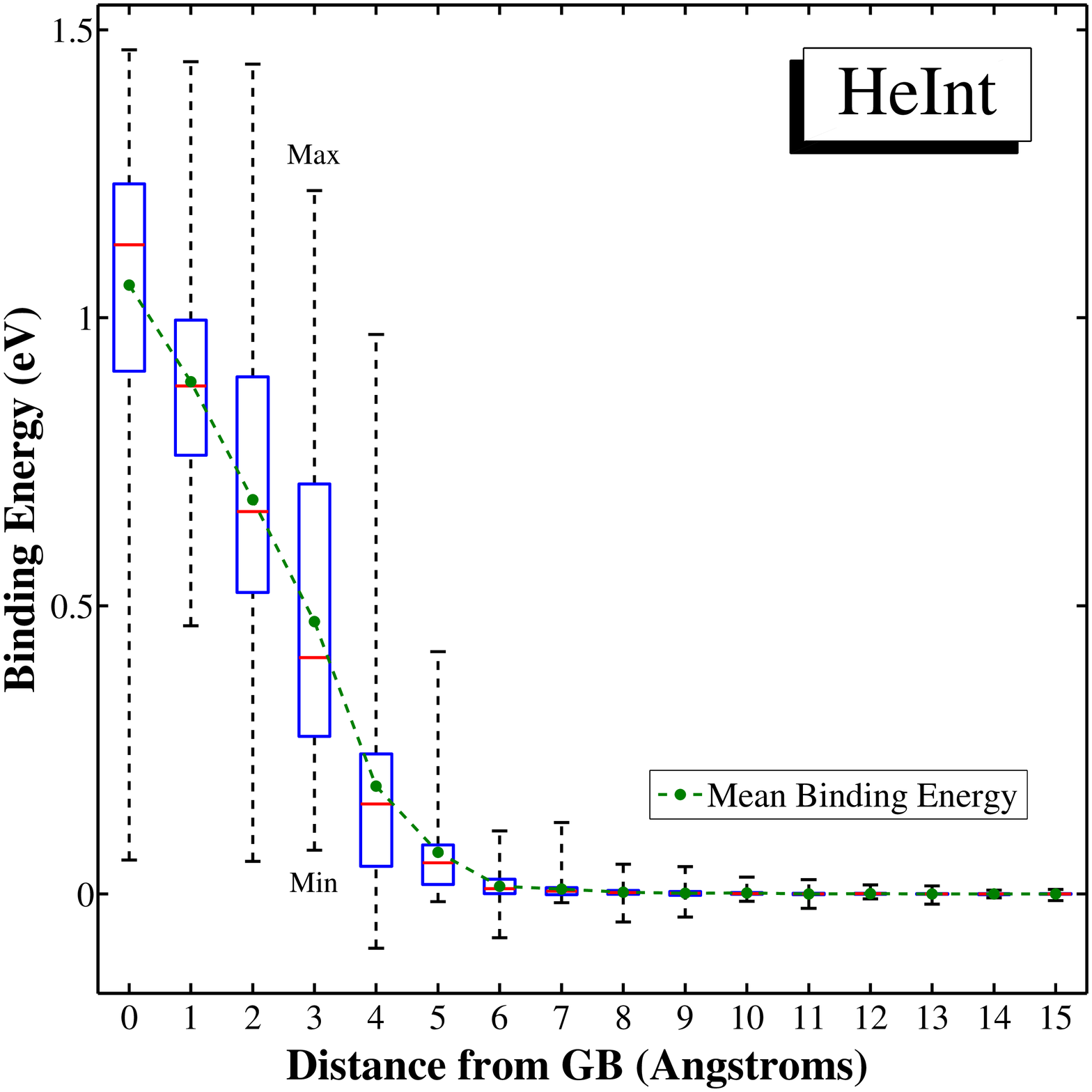} \\
	(a) & (b) \\
	 \includegraphics[width=0.45\columnwidth,angle=0]{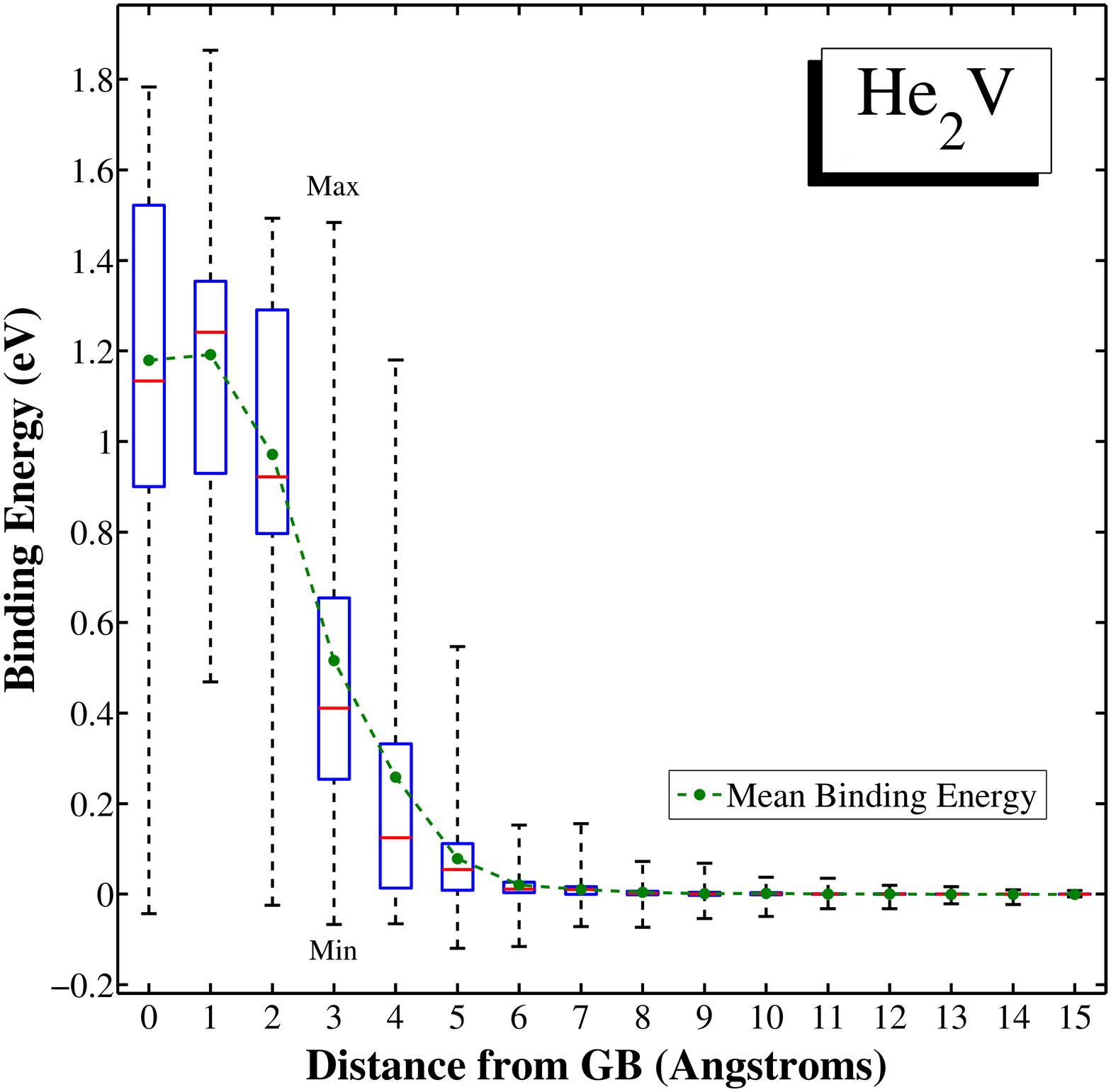} &
	 \includegraphics[width=0.45\columnwidth,angle=0]{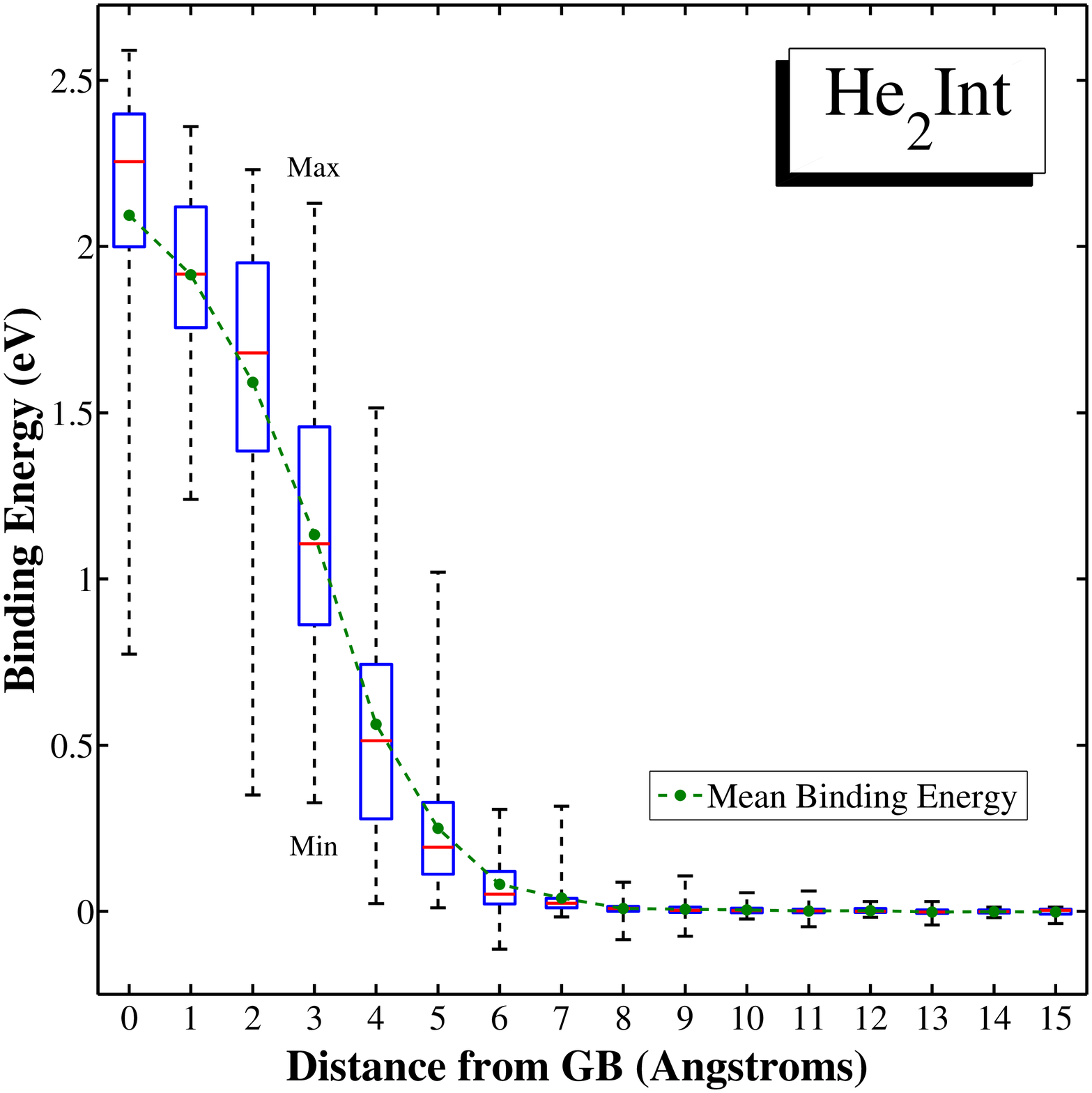} \\
	(c) & (d)
\end{tabular}
\caption{ \label{figure9} Boxplots of binding energy as a function of distance from the grain boundary for the nine representative GBs (excluding the $\Sigma3$\plane112) for various He defects: (a) HeV, (b) HeInt, (c) He$_2$V, and (d) He$_2$Int.  The data is divided into 1 \AA\ bins, and a boxplot is made for each bin. The red lines are medians, the blue box ends are the first and third quartiles, and the black whisker ends are minimum and maximum values.  The mean binding energy is also plotted in green.  }
\end{figure}

The change in binding energies of He defects as a function of distance from the GB center can also be analyzed by binning the energies and calculating the statistics associated with each bin (Figure \ref{figure9}).  Due to the symmetric nature of the GB formation and binding energies as a function of distance (e.g., Figure \ref{figure7}), the absolute value of the distance from the GB center was used to provide more data points for the statistical analysis.  Furthermore, the energies are split into 1 \AA\ bins to characterize the distributions and compute statistics for sites at a given distance from the GB.  For example, the 0 \AA\ bin would contain all binding energies for sites within $-0.5$ \AA\ to $+0.5$ \AA\ from the GB center and then several statistics are calculated from these binding energy distributions.  A boxplot (Figure \ref{figure9}) is used to represent the binding energy statistics in each bin, i.e., the minimum, 25\% percentile, median, 75\% percentile, and maximum binding energies.  In the boxplot, the red line in the box is the median while the top and bottom edges of the blue boxes represent the 25\% and 75\% quartiles. The whiskers extending from the boxes cover the remainder of the range of energies for each bin, and the ends of the whiskers denote the maximum and minimum values of the segregation energies for each bin.  The mean value of the segregation energies in each bin is also plotted in green.  Boxplots can be very useful for any asymmetry in the distribution of energies.

The box plots in Figure \ref{figure9} encompasses all the binding energy data for all four He defects from the nine representative GBs (excluding the $\Sigma3$\plane112).  The mean binding energy is lowest for sites close to the GB (0 and 1 \AA\ bins), as shown in Figure \ref{figure7}, and it approaches the normalized bulk value of zero as sites are located farther from the boundary.  The mean and median values of the binding energies trend together.  Similar to the trends found in Table \ref{table5}, there is a definite length scale associated with He defects binding to the grain boundary that is on the order of 5--6 \AA\ from the GB center.  For binding energies within the GB region, the distribution of binding energy is slightly skewed (for a symmetric distribution, the red line lies exactly in the middle of the box) with a large degree of variability, as can be seen from both the difference between the minimum and maximum values as well as the magnitude of the interquartile range (height of the boxes, denoting the binding energies associated with the 25\% and 75\% percentiles)).    At distances greater than 5--6 \AA, the boxes and extreme values are closely centered about the bulk value, which shows that the overwhelming majority of atomic sites display a binding energy similar to the bulk value.  

\subsection{Long Time Dynamics}

The long time dynamics of the two single He defects (interstitial He and HeV) at 300 K were also examined to determine whether the molecular statics simulations of formation and binding energies are effective figures or merit for predicting the GB sink efficiency for He.  While both defects were simulated, the HeV long time dynamics simulations were not included as the time scales for migration of the HeV defect in the absence of other defects (e.g., a vacancy) is so much larger than the interstitial He case, that the single crystal and both GBs essentially trap the HeV defect.  These simulations show that the He atom is not able to break away from the vacancy, either.  Even if the HeV defect separates into a vacancy and an interstitial He, the present simulations show that recombination occurs rather than migration of these defects away from each other.  Hence, the lower binding energy of substitutional He to the grain boundary compared to interstitial He is not as important as the energy barrier for migration of HeV. 

\begin{table}
\centering
\caption{\label{table7} Long time dynamics simulations of interstitial He at \SI{300}{K}}
\begin{footnotesize}
\begin{ruledtabular}
\begin{tabular}{lccc}
Defect & Bulk & $\Sigma3$\plane112  & $\Sigma11$\plane332 \\
\hline \\ [-1.5ex]
Total Time Simulated, $t$ &  \SI{18}{\pico\second} & \SI{12.3}{\nano\second} & 1.8 \si{\micro\second} \\
Formation Energy, $E_f$ &  4.38 eV & 4.22 eV  & 3.46 eV  \\
Migration Energy, $E_m$ & 0.058 eV & 0.058--0.160 eV  & 0.350 eV \\
Number of Events, $n_{events}$ & 69 & 110 & 30 \\
$n_{events}$ within 1$^{st}$ GB layers & N/A & (64/73, 87\%) & (21/30, 70\%) \\
Total Length, $l_{tot}$ & 87.0 \AA\ & 126.3 \AA\ & 74.6 \AA\ \\
End to End Length, $l_{end}$ & 10.5 \AA\ &  10.2 \AA\ &  9.5 \AA\ \\
\end{tabular}
\end{ruledtabular}
\end{footnotesize}
\end{table}

\begin{figure}[b!]
  \centering
	\includegraphics[width=0.8\columnwidth,angle=0]{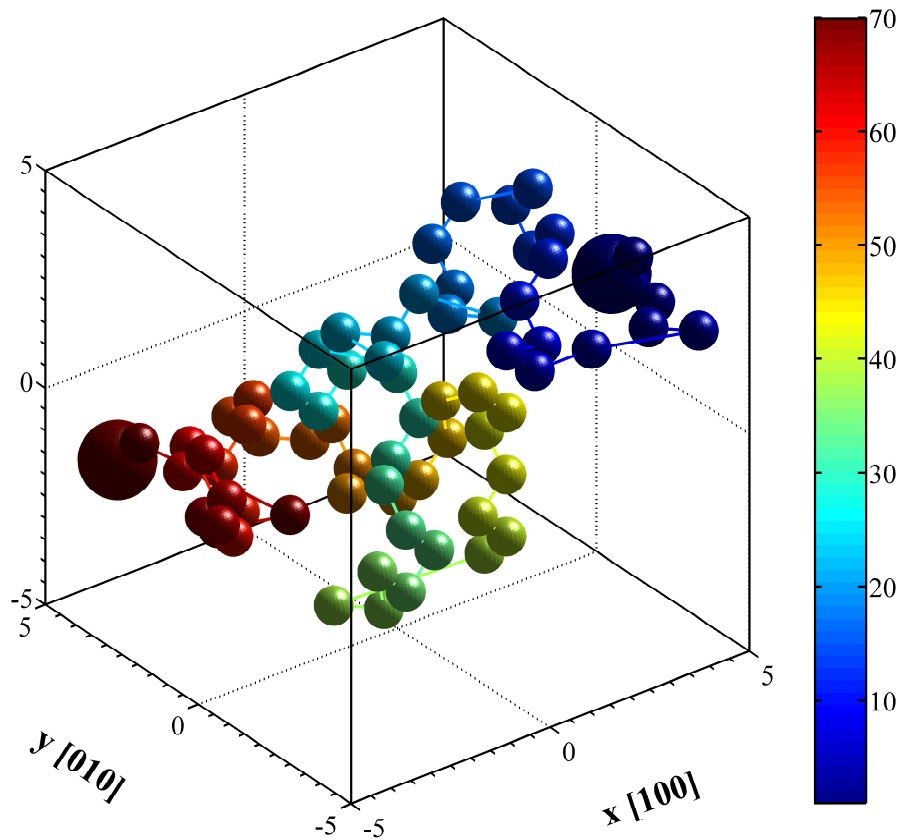} \\
\caption{\label{figureLTD1}Long time dynamics simulation of interstitial He in the BCC $\alpha$-Fe single crystal lattice.  The large atoms denote the beginning and ending positions of the interstitial He atom and the atom color represents the time evolution of the position of the He atom as it migrates through the lattice.}
\end{figure}

\begin{figure}[b!]
  \centering
	\begin{tabular}{cc}
	 \includegraphics[width=0.45\columnwidth,angle=0]{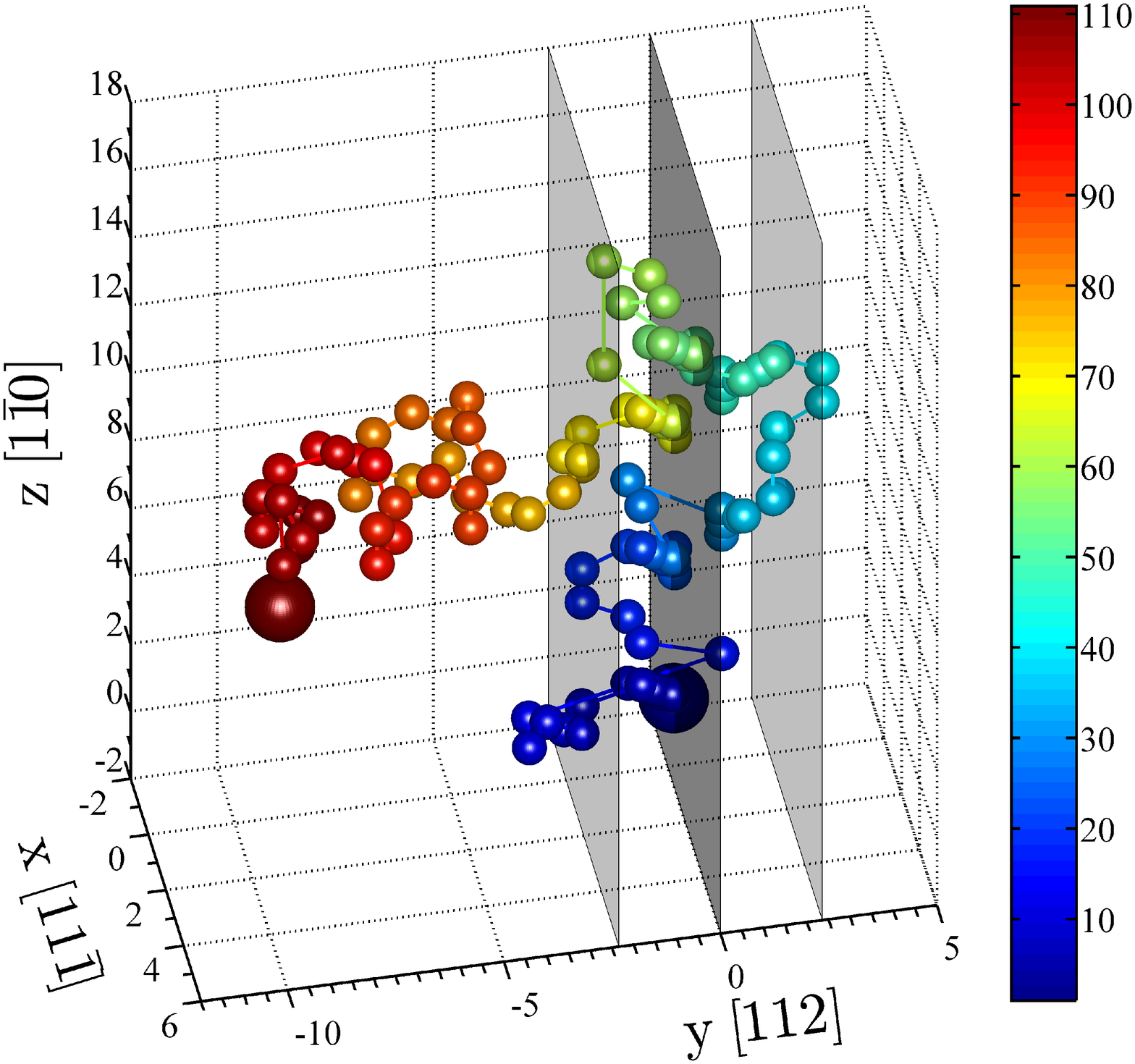} &
	 \includegraphics[width=0.45\columnwidth,angle=0]{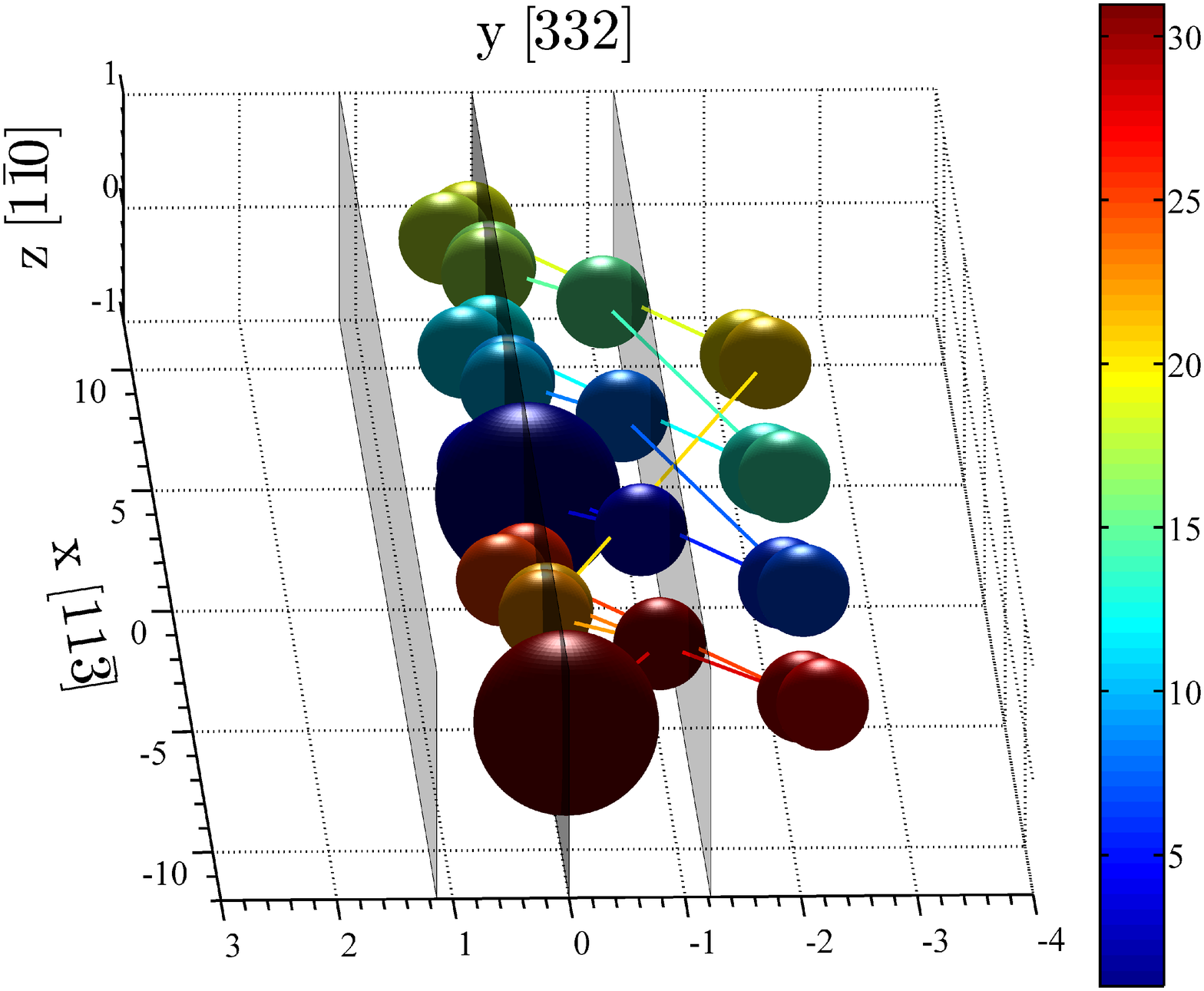} \\
	(a) & (b)
\end{tabular}
\caption{\label{figureLTD2}Long time dynamics simulation of interstitial He in the (a) $\Sigma3$\plane112 and (b) $\Sigma11$\plane332 GBs.  The large atoms denote the beginning and ending positions of the interstitial He atom and the atom color represents the time evolution of the position of the He atom as it migrates through the lattice.  The three planes represent the GB plane (darker, middle) and the first atomic layer on either side (lighter, adjacent to GB plane).}
\end{figure}

Table \ref{table7} shows the long time dynamics approach of Henkelman et al.~\cite{Hen2001,Ped2009} as it applies to interstitial He for the bulk single crystal, the $\Sigma3$\plane112 boundary, and the $\Sigma11$\plane332 GB.  For more details of the method, see Pedersen et al.~\cite{Ped2009} and the references therein.  For the GB simulations, the He defect was initially placed at the grain boundary and the formation energy $E_f$ listed in Table \ref{table7} corresponds to that site.  The migration energy $E_m$ was calculated for the successive event/movement of the He defect and the range is listed.  The long time dynamics method was simulated for the associated time $t$ and number of events $n_{events}$ in Table \ref{table7}.  The number of events within the 1$^{st}$ layers (for GBs), total length $l_{tot}$, and end to end length $l_{end}$ (net length traveled by interstitial He) is also listed.  The interstitial He migration through the bulk $\alpha$-Fe single crystal and the two GBs is shown in Figures \ref{figureLTD1} and \ref{figureLTD2}, respectively.  In these plots, the time evolution of the interstitial He migration is shown through coloring the atom according to the accepted events and the beginning/final position of the interstitial He is denoted by larger atoms (blue/red).  Additionally, for the grain boundaries in Figure \ref{figureLTD2}, the GB plane is shown (darker plane) as well as the first layer of atoms (adjacent lighter planes) to indicate whether the interstitial He can escape from the boundary.

The three different cases show different responses for interstitial He.  First, in the bulk single crystal, the migration of interstitial He is relatively easy over the simulated time of 18 ps.  In the grain boundary case, there are very different responses, though.  Interestingly, for the case of the interstitial He atom, the He is able to break away from the $\Sigma3$\plane112 boundary with a maximum migration energy of $E_m=0.160$ eV.  Since the long time simulation ran for a total time of 12.3 ns, the interstitial He broke away from the boundary in a time scale on the order of nanoseconds.  The number of events within the first layer of the GB was also included in Table \ref{table7}.  There were 73 total events prior to the interstitial He breaking away from the GB region and 64, or 87\%, of those events occured within 1 layer of the GB plane.  As can be seen in Figure \ref{figureLTD1}, on a number of cases, the interstitial He migrated outside of the 1$^{st}$ GB layer and then re-entered the GB region.  However, the low binding energy and small length scale of the $\Sigma3$\plane112 GB eventually allowed the interstitial He to break away over the time span of 12.3 ns.  On the other hand, in the $\Sigma11$\plane332 GB, the interstitial He predominantly migrates within the first couple of GB layers with a maximum migration energy of $E_m=0.350$ eV.  The long time dynamics simulations were simulated for 1.8 $\mu$s and only 4 times was the interstitial He able to migrate beyond the first layer.  In each of those times, the interstitial He migated back towards the GB plane and was effectively trapped at the boundary.  Interestingly, since the interstitial He atom cannot escape over the time span simulated in the present work, the net migration of interstitial He in the $\Sigma11$\plane332 GB is in the \dirf113 GB plane direction and not in the \dirf110 tilt direction (effectively, one-dimensional on the scale of $\mu$s).  These long time dynamic simulations validate that the anomalous behavior of the $\Sigma3$\plane112 boundary with respect to the molecular statics simulations (formation and binding energies) is associated with different dynamic behavior and sink efficiency as well, at least for the case of interstitial He.  However, these simulations also show that calculating the migration energies is also important for understanding the interaction between He defects and grain boundaries.  Interestingly, since the $\Sigma11$\plane332 has binding properties closest to the $\Sigma3$\plane112 (see Table \ref{table5}), this implies that interstitial He placed at the GB for all other boundaries studied within would be similarly trapped by the grain boundary.  

The results produced using the present Gao et al.~\cite{Gao2011} potential and the long time dynamics technique produce results that differ from earlier studies \cite{Gao2009}.  For instance, in the present work, an interstitial He atom was able to break away from the $\Sigma3$\plane112 boundary at 300 K, migrating three-dimensionally as in the bulk lattice upon breaking away from the boundary.  This is in contrast to earlier work that found that interstitial He primarily migrated one-dimensionally along the \dirf111 direction within the $\Sigma3$\plane112 GB at lower temperatures (600 K) and change to two-dimensional and three-dimensional migration at higher temperatures (800 K and 1000 K, respectively).  On the other hand, though, the $\Sigma11$\plane332 GB produced a zigzag behavior that predominantly lied within the grain boundary plane along the \dirf113 direction, similar to that observed by Gao et al.~\cite{Gao2009}.  This change in behavior for the $\Sigma3$\plane112 boundary may be attributed to the smaller migration energies for the current potential, which more accurately captures the energetics of He atoms within the Fe lattice \cite{Gao2011, Zha2013}.  This result is important as the $\Sigma3$\plane112 GB is the coherent twin in $\alpha$-Fe and has been experimentally observed at populations over 10 MRD in steels \cite{Bel2013a,Bel2013b}.  Moreover, this result shows the crucial role that the interatomic potential can have on nanoscale mechanisms and results.

The long time dynamics simulations show that the binding and formation energies for He defects can be important for understanding the physics of He diffusion and trapping by grain boundaries, which can be important for modeling the nucleation and growth of helium bubbles at grain boundaries.  The fact that the $\Sigma3$\plane112 grain boundary exhibits a low binding energy with He defects and our simulations indicate that interstitial He can break away from this boundary may have ramifications for grain boundary engineering of polycrystals, especially considering that this boundary is associated with twins within the microstructure.  The ability of process materials to increase the density of twins without appreciably affecting He trapping at these twin boundaries may be required for the extreme environment of future reactors. 

\section{Conclusions}

The formation/binding energetics and interaction length scales associated with the interaction between He atoms and grain boundaries in BCC $\alpha$-Fe was explored.  Ten different low $\Sigma$ grain boundaries from the \dirf100 and \dirf110 symmetric tilt grain boundary systems were used (Table \ref{table1}) along with an Fe--He interatomic potential fit to ab initio calculations \cite{Gao2011} (Table \ref{table2}).  In this work, we then calculated formation/binding energies for 1--2 He atoms in the substitutional and interstitial sites (HeV, HeInt, He$_2$V,  He$_2$Int) at all potential grain boundary sites within 15 \AA\ of the boundary (52826 simulations total).  For off-lattice He defects (HeInt, He$_2$V, He$_2$Int), 20 different random starting positions about each lattice site were selected.  The present results provide detailed information about the interaction energies and length scales of 1--2 He atoms with grain boundaries for the structures examined.  The following conclusions can be drawn from this work:
\begin{itemize}
	\item The local atomic structure and spatial location within the boundary affects the magnitude of the formation/binding energies for all four He defects (Figs.~\ref{figure2} and \ref{figure3}).  In general, grain boundary sites have much lower formation energies and higher binding energies than in the bulk, indicating an energetic driving force for He to segregate to grain boundaries.  This GB affected region visibly extends several planes from the GB center.  The maximum binding energy for the HeV defect within the 10 GBs is approximately 0.8 eV.  Furthermore, the $\Sigma3$\plane112 GB has significantly lower binding energies than all other GBs in this study, with the $\Sigma11$\plane332 GB having the second lowest binding energies of He defects.
	\item The binding energies for the twenty random starting positions for the off-lattice He defects were analyzed to provide information of the distribution of binding and formation energies at each grain boundary site (Figure \ref{figure3}).  The mean binding energy is more sensitive to local variations than the maximum binding energy and is more applicable to the energetic favorability of He defects than the standard deviation.  
	\item The relative binding energy behavior was examined with respect to grain boundary structure (e.g., Figure \ref{figure4}).  The binding energy behavior between the four defects is highly correlated in a positive sense ($R>0.9$), indicating that formation and binding energies of lower order defect types may be an adequate predictor of the formation energies of higher order defect types (e.g., using $E_f^{HeV}$ to accurately predict the mean $E_f^{He_2Int}$), which may be important for more expensive quantum mechanics simulations.
	\item Metrics for quantifying or classifying the local structure of each atom site were compared to the formation/binding energies of He defects.  Trends in per-atom metrics with each other and with the He defect energies were tabulated (Table \ref{table3} and \ref{table4}) and include: $E_{coh}$ positively correlated with vacancy binding energy $E_b^f$ ($R=0.89$), Voronoi volume $V_{Voro}$ positively correlated with hydrostatic stress $\sigma_H$ ($R=0.85$), centrosymmetry parameter and common neighbor analysis positively correlated with all He defect types ($R=0.76$ and $R=0.83$, respectively).  These correlations indicate that it may be possible to use metamodeling approaches to accurately predict formation and binding energies based on less computationally-expensive per-atom metrics.  
	\item The change in formation and binding energies as a function of spatial position (Figure \ref{figure7}) was used to identify a GB affected region and to assess a corresponding length scale, mean binding energy, and maximum binding energy for this region (Table \ref{table5}).  Table \ref{table5} indicates the $\Sigma3$\plane112 GB, and for some He defects the $\Sigma11$\plane332, does indeed stand out from the other GBs in the current study.   For the two-atom He defects, both the length scale and the binding energies are larger than for the single-atom He defects.  These values calculated using the Fe--He interatomic potential \cite{Gao2011} qualitatively agree with previous DFT calculations \cite{Zha2010i,Zha2013z} and both formation/binding energies agree with DFT within the calculated differences from Table \ref{table6}. These plots were additionally reduced via symmetry about the GB plane (Figure \ref{figure9}) to show the evolution of the binding energy distribution as a function of distance from the GB plane for the various defect types.
	\item Long time dynamics simulations \cite{Hen2001,Ped2009} for interstitial He placed at a site within the $\Sigma3$\plane112 and $\Sigma11$\plane332 GBs show that interstitial He can break away from the $\Sigma3$\plane112 GB on the order of nanoseconds at 300 K while interstitial He migrates predominantly one-dimensionally along the \dirf113 GB plane direction in the $\Sigma11$\plane332 GB and is effectively trapped by the boundary.
\end{itemize}

Atomistic simulations of this nature may ultimately help our understanding of how interface structure affects He diffusion to grain boundaries in polycrystalline steels.

\section*{Acknowledgments}

F.G.~and L.Y.~are grateful for the support by the US Department of Energy, Office of Fusion Energy Science, under Contract DE-AC06-76RLO 1830.  The authors would like to acknowledge the support and discussions with Xin Sun and Moe Khaleel at Pacific Northwest National Laboratory.  The authors would also like to acknowledge G.~Henkelman for aiding in implementing the long time dynamics algorithm into the molecular dynamics code utilized within.  Last, the authors would like to acknowledge Joanna Sun, high school student support by Alternate Sponsored Fellowship (ASF) at PNNL, for her contributions to this work.

\def\newblock{\hskip .11em plus .33em minus .07em}

\clearpage
\bibliographystyle{aip}


\end{document}